\documentclass[iop,apjl]{emulateapj}

\usepackage{amsmath}
\usepackage{mathrsfs}
\usepackage[varg]{txfonts}
\usepackage{bm}
\usepackage{cancel}
\usepackage{braket}

\usepackage[OT2,OT1]{fontenc}
\newcommand\cyr{
\renewcommand\rmdefault{wncyr}
\renewcommand\sfdefault{wncyss}
\renewcommand\encodingdefault{OT2}
\normalfont
\selectfont}
\DeclareTextFontCommand{\textcyr}{\cyr}

\renewcommand\today{2012 January 30}

\newcommand\Nu{\mathrm N}
\newcommand\Cr{\mathcal C}
\newcommand\rmd{\mathrm d}
\newcommand\rme{\mathrm e}
\newcommand\RLI[3]{{{}^+_{#1}\!{\textstyle\fint}_{\!#2}}^{#3}}
\newcommand\frd[3]{{{}^+_{#1}\partial_{\!#2}}^{#3}}

\newcommand\laplace[2]{\underset{#1\rightarrow#2}{\mathcal L}}
\newcommand\ilaplace[2]{\underset{#1\rightarrow#2}{\mathcal L^{-1}}}

\providecommand\abs[1]{\lvert{#1}\rvert}
\providecommand\norm[1]{\lVert{#1}\rVert}
\providecommand\flr[1]{\lfloor{#1}\rfloor}
\providecommand\clg[1]{\lceil{#1}\rceil}
\providecommand\frp[1]{\lbrace{#1}\rbrace}

\newtheorem{definition}{Definition}[section]
\newtheorem{theorem}[definition]{Theorem}
\newtheorem{corollary}[definition]{Corollary}
\newtheorem{lemma}[definition]{Lemma}
\begin{document}

\journalinfo{Preprint ver.\ \rm\today\ (\pageref{lastpage}pp)}
\shorttitle{\sc J.~An\hfill}\shortauthors{\hfill\sc J.~An}
\title{Fractional calculus, completely monotonic functions,
a generalized Mittag-Leffler function and
phase-space consistency of separable augmented densities}
\author{Jin H. An}
\slugcomment{\tt jinan@nao.cas.cn}

\begin{abstract}\noindent
Under the separability assumption on the augmented density,
a distribution function can be always constructed for
a spherical population with the specified density and anisotropy
profile. Then, a question arises, under what conditions
the distribution constructed as such
is non-negative everywhere in the entire accessible subvolume
of the phase-space. We rediscover necessary conditions
on the augmented density expressed with fractional calculus.
The condition on the radius part $R(r^2)$ -- whose logarithmic
derivative is the anisotropy parameter -- is equivalent to
$w^{-1}R(w^{-1})$ being a completely monotonic function whereas
the condition on the potential part is stated
as its derivative up to the order not greater than $\frac32-\beta_0$
being non-negative (where $\beta_0$ is the central limiting value
for the anisotropy parameter). 
We also derive the set of sufficient conditions on
the separable augmented density for the non-negativity
of the distribution, which generalizes the condition
derived for the generalized \citeauthor{Cu91} system by \citet{CM10}
to arbitrary separable systems. This is applied for the case
when the anisotropy is parameterized by a monotonic function of
the radius of \citet{BvH07}. The resulting criteria are
found based on the complete monotonicity of generalized
Mittag-Leffler functions.
\end{abstract}

\section{Models for spherical dynamical system}
\subsection{Distribution function}

Suppose that $\mathcal F(\bm r;\bm v|t)$
is a phase-space distribution so that
\[\int_S\mathcal F\rmd^3\!\bm r\,\rmd^3\!\bm v\]
is the number of tracers in any measurable phase-space volume $S$
at time $t$. Here $\bm r$ is the position vector in the configuration
space and $\bm v=\dot{\bm r}$ is the velocity. We only consider the
system in equilibrium and thus the distribution must be time-independent.
The distribution of a spherically symmetric population in a steady state
is also invariant under transforms in SO(3) so that
$\mathcal F(\bm r;\bm v|t)=\mathcal F(r;v_r,\norm{\bm v_\mathrm t})$
where $r=\norm{\bm r}$ and $\hat{\bm r}=\bm r/r$ are
the radial distance and unit vector while
$v_r=\bm{v\cdot}\hat{\bm r}$ and
$\bm v_\mathrm t=\bm v-v_r\hat{\bm r}$
are the radial and tangential velocities.
If we adopt the canonical spherical polar coordinate $(r,\theta,\phi)$,
they are given by
\[v=\norm{\bm v}=(v_r^2+v_\mathrm t^2)^\frac12\,;\quad
v_\mathrm t=\norm{\bm v_\mathrm t}=(v_\theta^2+v_\phi^2)^\frac12,\]
where $(v_r,v_\theta,v_\phi)=(\dot r,r\dot\theta,r\dot\phi\sin\theta)$
are the velocity components projected onto the associated orthonormal basis.

In order for the distribution to be indeed time-independent,
it must be invariant under dynamic evolutions of tracers, that is,
the distribution is a time-independent
solution to the Boltzmann\footnote{Ludwig Eduard Boltzmann (1844-1906)}
transport equation. For typical stellar dynamical applications,
the trajectory of each tracer is
its orbit under the external potential, which may
or may not be self-consistently generated by the tracer population.
The transport equation for this case results in the \emph{collisionless}
Boltzmann equation (CBE), whose solution is completely characterized by
the theorem due to J. Jeans\footnote{Sir James Hopwood Jeans (1877-1946)}.
The \citeauthor{Je15} theorem indicates that
if the given time-independent spherically-symmetric distribution
function (df) is a solution to CBE with a \emph{generic} static
spherical potential $\Phi(r)$, it must be in the form of
\[\mathcal F=\mathcal F(\mathcal E,L^2)\]
where
\[\textstyle\mathcal E=\Psi(r)-\frac12{v^2}\,;\quad
L=\norm{\bm L}=rv_\mathrm t,\]
are the two isotropic isolating integrals admitted by
all such potentials, namely, the specific binding energy and
the magnitude of the specific angular momentum, respectively. Here,
\[\Psi(r)\equiv\begin{cases}
\Phi(r_\mathrm{out})-\Phi(r)&\text{if $r_\mathrm{out}$ is finite}\\
\Phi(\infty)-\Phi(r)&
\text{if $r_\mathrm{out}=\infty$ and $\abs{\Phi(\infty)}<\infty$}\\
-\Phi(r)&\text{if $r_\mathrm{out}=\infty$ and
$\Phi(\infty)\rightarrow\infty$}\\
\end{cases}\]
is the \emph{relative potential}
with respect to the boundary $r_\mathrm{out}$.
The system not bounded by a finite boundary radius
is represented by $r_\mathrm{out}=\infty$ with
$\Phi(\infty)=\lim_{r\rightarrow\infty}\Phi(r)$.
If $r_\mathrm{out}$ or $\Phi(\infty)$ is finite, then
$\mathcal F(\mathcal E<0,L^2)=0$ because by definition
$\mathcal E\ge0$ for all tracers bound to the system
(and bounded by $r\le r_\mathrm{out}$).

\subsection{Augmented densities of a spherical system}

Integrating the spherical two-integral df
$\mathcal F(\mathcal E,L^2)$ over the velocity space results in
a bivariate function of $\Psi$ and $r^2$,
\begin{equation}\label{eq:ad}
\Nu(\Psi,r^2)\equiv\iiint\!\rmd^3\!\bm v\,
\mathcal F\bigl(\mathcal E=\Psi-\tfrac12v^2,L^2=r^2v_\mathrm t^2\bigr),
\end{equation}
which is referred to as the \emph{augmented density} (AD).
The integral here is formally over the whole velocity
subspace, but if $r_\mathrm{out}$ or $\Phi(\infty)$ is finite, it is
essentially within the sphere $v^2\le2\Psi$ since
$\mathcal F(\mathcal E<0,L^2)=0$ for these cases. With $\Psi(r)$ specified,
the AD yields the local density $\nu(r)$ via \[\nu(r)=\Nu[\Psi(r),r^2].\]
Similarly, the \emph{augmented moment} functions are given by
\begin{subequations}
\begin{multline}
m_{k,n}(\Psi,r^2)\equiv\iiint\!\rmd^3\!\bm v\,v_r^{2k}v_\mathrm t^{2n}
\mathcal F\bigl(\mathcal E=\Psi-\tfrac12v^2,L^2=r^2v_\mathrm t^2\bigr)
\\=4\pi\!
\iint\limits_{\substack{(v^2\le2\Psi)\\v_r\ge0,v_\mathrm t\ge0}}\!
\rmd v_r\,\rmd v_\mathrm t\,v_r^{2k}v_\mathrm t^{2n+1}
\mathcal F\Bigl(\Psi-\frac{v_r^2+v_\mathrm t^2}2,r^2v_\mathrm t^2\Bigr).
\end{multline}
Changing the integration variables to $(\mathcal E,L^2)$,
these are represented to be a set of integral transformations of the df,
\begin{multline}\label{eq:dist}
m_{k,n}=\frac{2\pi}{r^{2n+2}}\!
\iint_\mathcal T\!\rmd\mathcal E\,\rmd L^2
\mathcal K^{k-\frac12}L^{2n}\mathcal F(\mathcal E,L^2)
\\=\frac{2\pi}{r^{2n+2}}\!
\iint_{\mathcal E\ge\mathcal E_0,L^2\ge0}\!
\rmd\mathcal E\,\rmd L^2\Theta(\mathcal K)\,
\abs{\mathcal K}^{k-\frac12}L^{2n}\mathcal F(\mathcal E,L^2).
\end{multline}
Here $\Theta(x)$ is the Heaviside\footnote{Oliver Heaviside (1850-1925)}
unit-step function and
\[\mathcal E_0\equiv\begin{cases}
0&\text{if $r_\mathrm{out}$ or $\Phi(\infty)$ is finite}\\
-\infty&\text{if
$\lim_{r\rightarrow\infty}\Psi(r)=-\Phi(\infty)\rightarrow-\infty$}
\end{cases}\]
is the lower bound of the binding energy.
The transform kernel and the domain in $(\mathcal E,L^2)$
space over which the integral is performed are given by
\begin{align*}&\mathcal K(\mathcal E,L^2;\Psi,r^2)\equiv
2(\Psi-\mathcal E)-r^{-2}L^2,
\\&\mathcal T\equiv\set{(\mathcal E,L^2)|
\mathcal E\ge\mathcal E_0,L^2\ge0,\mathcal K\ge0}.\end{align*}
Note $\mathcal K$ is $v_r^2$ expressed as a function of
$4$-tuple $(\mathcal E,L^2;\Psi,r^2)$.
\end{subequations}

\section{Mathematical preliminary}
\subsection{Fractional calculus}

\begin{definition}
The Riemann\footnote{Georg Friedrich Bernhard Riemann
(1826-1866)}-Liouville\footnote{Joseph Lioville (1809-1882)}
integral operator of arbitrary non-negative real order $\lambda\ge0$
is given by
\begin{equation}\label{eq:mint}
\RLI ax\lambda f
\equiv\begin{cases}{\displaystyle
\frac1{\Gamma(\lambda)}\!\int_a^x\!(x-y)^{\lambda-1}f(y)\,\rmd y
}&(\lambda>0)\\f(x)&(\lambda=0)\end{cases},
\end{equation}
where $\Gamma(x)$ is the gamma function.
\end{definition}
This is a trivial generalization
of the Cauchy\footnote{Augustin-Louis Cauchy (1789-1857)} formula
for repeated integrations. For $0<\lambda<1$,
this is also recognized as the generalized
Abel\footnote{Niels Henrik Abel (1802-1829)} transform with
the classical case corresponding to the $\lambda=\frac12$ case.
We also define
\begin{definition}
the fractional derivative for $\lambda\ge0$
such that
\begin{multline}\label{eq:frd}
\frd ax\lambda f
\equiv\frac{\rmd^{\clg\lambda}}{\rmd x^{\clg\lambda}}
\RLI ax{\clg\lambda-\lambda}f
\\=\begin{cases}{\displaystyle
\frac1{\Gamma(1-\frp\lambda)}
\frac{\rmd^{\clg\lambda}}{\rmd x^{\clg\lambda}}\!
\int_a^x\!\frac{f(y)\,\rmd y}{(x-y)^{\frp\lambda}}
}&(0<\frp\lambda<1)\smallskip\\
\dfrac{\rmd^\lambda f(y)}{\rmd x^\lambda}\biggr\rvert_{y=x}=f^{(\lambda)}(x)
&(\frp\lambda=0)\end{cases}
\end{multline}
where $\clg\lambda$, $\flr\lambda$, and
$\frp\lambda=\lambda-\flr\lambda$ are the integer ceiling,
the integer floor and the fractional part of $\lambda$, respectively.
\end{definition}
Note equation (\ref{eq:frd}) is a generalization of
the differentiation for positive real order
as is equation (\ref{eq:mint}) of the integration.
These definitions extend to include a negative index using
\begin{definition}
for arbitrary real $\lambda$
\begin{equation}\label{eq:mintn}
\RLI ax{-\lambda}f=\frd ax\lambda f
\quad\text{and vice versa}.
\end{equation}
\end{definition}

The basic composite rule for the Riemann-Liouville operators is that,
for any pair of non-negative reals $\lambda$ and $\xi$,
\begin{equation}\label{eq:intcomp}
\RLI ax\xi\Bigl(\RLI ax\lambda f\Bigr)=\RLI ax{\xi+\lambda}f,
\end{equation}
which may be shown by direct calculations
using the Fubini\footnote{Guido Fubini (1879-1943)} theorem and
the Euler\footnote{Leonhard Euler (1707-1783)} integral of
the first kind for the beta function, that is,
\[\begin{split}
\int_a^x&\rmd y\,(x-y)^{\xi-1}\!\int_a^y\!\rmd w\,(y-w)^{\lambda-1}f(w)
\\&=\int_a^x\!\rmd w\,f(w)\!\int_w^x\!\rmd y\,(x-y)^{\xi-1}(y-w)^{\lambda-1}
\\&=\int_a^x\!\rmd w\,f(w)\,(x-w)^{\xi+\lambda-1}
\!\int_0^1\!\rmd t\,(1-t)^{\xi-1}t^{\lambda-1}.
\end{split}\]
Next for any real $\lambda$ and a non-negative integer $n$
\begin{equation}
\frac\rmd{\rmd x}\RLI ax\lambda f=\RLI ax{\lambda-1}f\,;\quad
\frac{\rmd^n}{\rmd x^n}\RLI ax\lambda f=\RLI ax{\lambda-n}f.
\end{equation}
Here the latter follows the former ($n=1$) by means of induction.
The $n=1$ case is proven by direct
differentiation of equation (\ref{eq:mint}) for $\lambda>1$ and
the fundamental theorem of calculus for $\lambda=1$ while
the same case with $\lambda<1$ is essentially trivial from
the definitions of fractional derivatives in equations (\ref{eq:frd})
and (\ref{eq:mintn}). Together they also indicate that
\begin{equation}\label{eq:difcomb0}
\frd ax\xi\Bigl(\RLI ax\lambda f\Bigr)=\begin{cases}
\RLI ax{\lambda-\xi}f&(\xi\le\lambda)\\
\frd ax{\xi-\lambda}f&(\xi\ge\lambda)\end{cases},
\end{equation}
for non-negative reals $\lambda,\xi\ge0$ and arbitrary function $f(x)$,
provided that all the integrals in their respective definitions
absolutely converge.
Next we observe for $\lambda\ge0$ that
\begin{subequations}\begin{equation}
\RLI ax{\lambda+1}f^\prime=\RLI ax\lambda f
-\frac{(x-a)^\lambda f(a)}{\Gamma(\lambda+1)},
\end{equation}
thanks to the fundamental theorem of calculus ($\lambda=0$)
and integration by part. By means of induction, this generalizes to
\begin{equation}
\RLI ax{\lambda+n}f^{(n)}=\RLI ax\lambda f
-\sum_{k=0}^{n-1}\frac{(x-a)^{\lambda+k}f^{(k)}(a)}{\Gamma(\lambda+k+1)},
\end{equation}\end{subequations}
where $n$ is any non-negative integer, and we also find that
\begin{equation}\label{eq:ibpr}
\frac{\rmd^n}{\rmd x^n}\RLI ax\lambda f
=\RLI ax\lambda f^{(n)}+\sum_{k=1}^n
\frac{(x-a)^{\xi-k}f^{(n-k)}(a)}{\Gamma(1+\lambda-k)}
\end{equation}
for $\lambda\ge0$ and any non-negative integer $n$.
The last implies that fractional derivatives
in equation (\ref{eq:frd}) are alternatively given by
\begin{equation}\label{eq:frda}
\frd ax\lambda f
=\frac{\rmd^{\clg\lambda-n}}{\rmd x^{\clg\lambda-n}}
\RLI ax{\clg\lambda-\lambda}f^{(n)}+\sum_{k=0}^{n-1}
\frac{(x-a)^{k-\lambda}f^{(k)}(a)}{\Gamma(1+k-\lambda)}
\end{equation}
where $\lambda>0$ and $n=0,1,\dotsc,\clg\lambda$.

Using these and equation (\ref{eq:ibpr}), we can also derive that
\begin{equation}\label{eq:difcomb}\begin{split}
\RLI ax\xi\frd ax\lambda f&=\RLI ax{\xi-\lambda}f
-\sum_{k=1}^{\flr\lambda}C^+_{\xi,k}\,
\frd ax{\lambda-k}\!f(a)\,(x-a)^{\xi-k}
\\\frd ax\xi\frd ax\lambda f&=\frd ax{\xi+\lambda}f
-\sum_{k=1}^{\flr\lambda}C^-_{\xi,k}
\frac{\frd ax{\lambda-k}\!f(a)}{(x-a)^{k+\xi}}
\end{split}\end{equation}
for non-negative reals $\lambda,\xi\ge0$ and arbitrary function $f(x)$,
provided again that all the integrals in their respective definitions
absolutely converge. Here $C^\pm_{\xi,k}$ are given by
\[C^\pm_{\xi,k}=\frac1{\Gamma(1\pm\xi-k)}=\begin{cases}
\dfrac{(\xi)_k^-}{\Gamma(1+\xi)}&\text{($+$ case)}\smallskip\\
\dfrac{(-1)^{\flr\xi+k}(\delta)_{\flr\xi+k}^+}{\Gamma(1-\delta)}
&\text{($-$ case)}\end{cases}\]
where $0\le\delta=\xi-\flr\xi<1$ is the fractional part of $\xi$, and
\[\textstyle
(a)_n^-={\textstyle\prod_{j=1}^n(a+1-j)}\,;\quad
(a)_n^+={\textstyle\prod_{j=1}^n(a-1+j)}\]
are the Pochhammer\footnote{Leo August Pochhammer (1841-1920)} symbol.
The \emph{falling} product $(a)_n^-$ follows
the combinatorist's convention whereas the \emph{rising} one
$(a)_n^+$ does the analyst's. Note these are related to each other,
\[(-a)_n^-=(-1)^n(a)_n^+\,;\quad
(a)_n^-=(a-n+1)_n^+\]
and also to the gamma functions,
\[(a)_n^+=\frac{\Gamma(a+n)}{\Gamma(a)}\,;\quad
(a)_n^-=\frac{\Gamma(1+a)}{\Gamma(1+a-n)}.\]
The last may be used to generalize the Pochhammer symbol for
non-integer $n$.
Together equations (\ref{eq:difcomb0}) and (\ref{eq:difcomb})
provide the generalization of equation (\ref{eq:intcomp})
for any pair of reals $\xi$ and $\lambda$.

The simplest specific result of fractional calculus would be
\begin{lemma}\label{lem:frd}
for real $\lambda$ and $\alpha>0$,
\begin{equation}
\RLI 0x\lambda x^{\alpha-1}
=\frac{\Gamma(\alpha)\,x^{\alpha+\lambda-1}}{\Gamma(\alpha+\lambda)}.
\end{equation}
\end{lemma}
This is formally a generalization of the result, namely
\begin{equation}
\frac{\rmd^n x^\alpha}{\rmd x^n}
=(\alpha)_n^-x^{\alpha-n}
\qquad(n=0,1,\dotsc)
\end{equation}
although the last is in fact valid for any $\alpha$.

We formalize an obvious but important fact, namely
\begin{lemma}\label{lem:pos}
for $\lambda>0$ and $x>a$, if $f\ge0$ in $[a,x]$, then
$\RLI ax\lambda f(x)\ge0$. Moreover $\RLI ax\lambda f\ne0$
provided that the support of $f$ in $(a,x)$ has non-zero measure.
\end{lemma}
Next, if $a$ is finite, then for $\xi>0$
\[\RLI ax\xi f
=\frac{(x-a)^\xi}{\Gamma(\xi)}\!
\int_0^1\!\rmd t\,t^{\xi-1}f\bigl[x-(x-a)t\bigr],\]
while for $0<\lambda=-\xi<1$ and $n=1$,
equation (\ref{eq:frda}) results in
\[\RLI ax\xi f=\frd ax\lambda f
=\frac{f(a)}{\Gamma(1-\lambda)}\,(x-a)^{-\lambda}
+\RLI ax{1-\lambda}f'.\]
It then follows that
\begin{lemma}
for $a\ne\pm\infty$,
\begin{equation}\label{eq:limxif}
\lim_{x\rightarrow a^+}\frac{\RLI ax\xi f(x)}{(x-a)^\xi}
=\frac{f(a)}{\Gamma(\xi+1)},
\end{equation}
\end{lemma}
which is valid for $\xi\ge0$ if $f(x)$ is right-continuous at $x=a$
or for $\xi\ge-1$ if $f(x)$ is right-differentiable at $x=a$.
Equation (\ref{eq:limxif}) for $\xi=0$ is equivalent to
the definition of the right-continuity while for $\xi=-1$,
it becomes $\lim_{x\rightarrow a^+}(x-a)f'(x)=0$
which holds if $f'(a)$ is finite.
Equation (\ref{eq:limxif}) implies that
\begin{corollary}\label{eq:if0}
if $f(x)$ is right-continuous at $x=a$ and $f(a)$ is finite,
then $\RLI ax\lambda f(a)=0$ for $\lambda>0$.
\end{corollary}

Next we examine the behaviors of fractional calculus operators
under the Laplace\footnote{Pierre-Simon 
Laplace (1749-1827)} transform.
For this, we first note a general property of the Laplace transform
of the derivative,
\begin{equation}\label{eq:lapderiv}
s^{n+1}\laplace xs[f(x)]
=\laplace xs[f^{(n+1)}(x)]
+{\textstyle\sum_{j=0}^ns^jf^{(n-j)}(0)},
\end{equation}
which is valid 
given that $\lim_{x\rightarrow\infty}\rme^{-sx}f^{(n)}(x)=0$ for
sufficiently large $s$ (which is required for the Laplace transform
to converge). Equation (\ref{eq:lapderiv}) is proven
via integration by part,
\[\int_0^\infty\!\rmd x\,\rme^{-sx}\frac{\rmd f(x)}{\rmd x}
=-f(0)+s\!\int_0^\infty\!\rmd x\,\rme^{-sx}f(x)\]
for $n=0$ and the induction completes its proof for any
non-negative integer. In order to generalize
equation (\ref{eq:lapderiv}) to include the fractional derivative,
we next consider for $\lambda\ge0$
%
\[\begin{split}
\int_0^\infty\!\rmd x\,&\rme^{-sx}\!
\int_0^x\!\rmd y\,(x-y)^{\lambda-1}f(y)
\\&=\int_0^\infty\!\rmd y\,f(y)\!
\int_y^\infty\!\rmd x\,(x-y)^{\lambda-1}\rme^{-sx}
\\&=\int_0^\infty\!\rmd y\,f(y)\,\frac{\rme^{-sy}}{s^\lambda}\!
\int_0^\infty\!\rmd u\,u^{\lambda-1}\rme^{-su}.
\end{split}\]
With the Euler integral of the second kind for the gamma function,
we find that
\begin{equation}\label{eq:lapab}
s^\lambda\laplace xs\Bigl[\RLI 0x\lambda f(x)\Bigr]
=\laplace xs[f(x)].
\end{equation}
The Laplace transform of an arbitrary real-order derivative
is then found by combining equations (\ref{eq:lapderiv})
and (\ref{eq:lapab}).

\subsection{Post--Widder formula}

\begin{theorem}[\emph{Post}--Widder]
If $\phi(t)$ is continuous for $t\ge0$ and there exist real
${}^\exists A>0$ and ${}^\exists b>0$ such that
\[\rme^{-bt}\abs{\phi(t)}\le A
\quad\text{for ${}^\forall t>0$,}\]
then the Laplace transform
\begin{equation}\label{eq:lap}
f(x)=\laplace tx[\phi(t)]
\equiv\int_0^\infty\!\rmd t\,\rme^{-xt}\phi(t).
\end{equation}
converges and is infinitely differentiable for $x>b$. Moreover,
$\phi(t)$ for $t>0$ may be inverted from $f(x)$
using the differential inversion formula \citep{Po30,Wi41},
\begin{equation}\label{eq:pinv}
\phi(t)=\ilaplace xt[f(x)]
=\lim_{n\rightarrow\infty}
\frac{(-1)^n}{n!}\,\Bigl(\frac nt\Bigr)^{n+1}
f^{(n)}\Bigl(\frac nt\Bigr).
\end{equation}
\end{theorem}
In literature, the last formula is typically named after
E. Post\footnote{Emil Leon Post (1897-1954)} or together with
D. Widder\footnote{David Vernon Widder (1898-1990)}.
A rigorous proof, which is beyond the scope of this paper,
may be found in a standard text on the Laplace transform.
However its heuristic justifications abound and are easy to observe.
For instance, direct calculations using equation (\ref{eq:lap})
indicate that
\[f^{(n)}(x)
=(-1)^n\!\int_0^\infty\!\rmd t\,t^n\rme^{-xt}\phi(t)
=\frac{(-1)^n}{x^{n+1}}\!\int_0^\infty\!\rmd s\,s^n\rme^{-s}
\phi\Bigl(\frac sx\Bigr).\]
and thus we find that
\[\frac{(-1)^n}{n!}\,\Bigl(\frac nt\Bigr)^{n+1}
f^{(n)}\Bigl(\frac nt\Bigr)
=\int_0^\infty\!\rmd s\,P(s;n)
\,\phi\Bigl(\frac snt\Bigr).\]
where
\[P(s;n)\equiv\frac{s^n}{n!}\,\rme^{-s}\] is the probability
density of the Poisson\footnote{Sim\'eon Denis Poisson (1781-1840)}
distribution with a mean of $\bar s=n$.
It follows that as $n\rightarrow\infty$,
the relative dispersion decreases and so
$\overline{\phi(st/n)}\rightarrow\phi(\bar st/n)=\phi(t)$, which
results in the Post--Widder formula.
Note however that the convergence of equation (\ref{eq:pinv}) by itself
does not necessarily imply that $f(x)$ is the Laplace transformation
of $\phi(t)$, which is rather a part of the condition for the formula
to be valid.

\subsection{Completely monotonic functions}

\begin{definition}\label{def:cm}
A smooth function $f(x)$ of $x>0$ is said to be
completely monotonic (cm henceforth) if and only if
\begin{equation}
(-1)^nf^{(n)}(x)\ge0
\qquad(x>0,\,n=0,1,2,\dotsc).
\end{equation}
\end{definition}
The definition extends to $x\ge0$ if $f(x)$ is right-continuous at $x=0$.
Some basic properties of cm functions are:
\begin{lemma}\label{lem:cm}
Let $f$ and $g$ be cm. Then,
\begin{enumerate}
\item $(-1)^nf^{(n)}$ for any non-negative integer $n$ is cm.
\item If $F\ge0$ in $(0,\infty)$ and $f=-F'$, then $F$ is cm.
\item $\int_x^\infty\!f(y)\,\rmd y$ is a cm function of $x$
if it converges.
\item $af+bg$ where $a$ and $b$ are non-negative constants is cm.
\item $f\cdot g$ is cm.
\item If $F>0$ in $(0,\infty)$ and $f=F'$, then $g\circ F$ is cm.
\item $\exp(f)$ is cm.
\end{enumerate}
\end{lemma}
Items {\it1} and {\it2} are essentially trivial from Defintion \ref{def:cm}
and item {\it3} is simply a particular case of item {\it2}.
Item \textit4 follows the linearity of differentiations
while item \textit5 is shown using
the Leibniz\footnote{Gottfried Wilhelm 
Leibniz (1646-1716)} rule,
that is, (here $\binom nk$ is the binomial coefficient)
\begin{equation}\label{eq:leibniz}
(-1)^n\frac{\rmd^n(f\cdot g)}{\rmd x^n}=\sum_{k=0}^n\binom nk\,
(-1)^k\frac{\rmd^kf}{\rmd x^k}\,
(-1)^{n-k}\frac{\rmd^{n-k}g}{\rmd x^{n-k}}.
\end{equation}
The last two may be shown using
the Fa\`a di Bruno\footnote{Francesco Fa\`a di Bruno (1825-1888)} formula
(i.e., the generalized chain rule),
\begin{equation}\label{eq:fdb}
(g\circ F)^{(n)}(t)
={\textstyle\sum_{k=0}^ng^{(k)}\bigl[F(t)\bigr]\cdot
B_{n,k}\bigl[f(t),f'(t),\dotsc,f^{(n-k)}(t)\bigr]}
\end{equation}
where $F'(t)=f(t)$ and $B_{n,k}$ is
the Bell\footnote{Eric Temple Bell (1883-1960)} polynomial, that is,
\[B_{n,k}(x_0,\dots,x_{n-k})
=\sideset{}{'}\sum_{(j_0,j_1,\dotsc)}
\frac{n!}{j_0!j_1!\dotsm}
\left(\frac{x_0}{1!}\right)^{j_0}
\left(\frac{x_1}{2!}\right)^{j_1}\dotsm.\]
Here the summation is over all sequences
$(j_0,j_1,\dotsc)$ of \emph{non-negative integers} constrained such that
\[\textstyle\sum_{m=0}j_m=k\,;\quad\sum_{m=0}(m+1)j_m=n.\]
Note then
\[\textstyle\sum_{m=0}mj_m=n-k\] and thus $j_m\ge0$ indicates that
$j_m=0$ for $m>n-k$ (n.b., if otherwise, $\sum_{m=0}mj_m>n-k$, which is
contradictory). The property \textit6 follows this because
\[\textstyle n-k-\sum_{m=0}j_{2m+1}=2\sum_{m=0}m\,(j_{2m}+j_{2m+1})\]
is even. That is to say,
if $f$ is cm, the parity of the Bell polynomial in
equation (\ref{eq:fdb}) is $(-1)^{n-k}$, and thus,
given that $g$ is also cm, the parity of every term in the sum
on the right-hand side of equation (\ref{eq:fdb}) is $(-1)^n$. 
Equation (\ref{eq:fdb}) also indicates that
\begin{equation}\label{eq:expf}
\frac{\rmd^n\exp[f(t)]}{\rmd t^n}
=\exp[f(t)]\cdot
B_n\bigl[f'(t),f''(t),\dotsc,f^{(n-k+1)}(t)\bigr]
\end{equation}
where $B_n$ is the $n$-th complete Bell polynomial, that is,
\[B_n(x_1,\dots,x_n)
={\textstyle\sum_{k=1}^nB_{n,k}(x_0,\dots,x_{n-k})}.\]
Note \[\textstyle n-\sum_{m=0}j_{2m}=2\sum_{m=0}m\,(j_{2m-1}+j_{2m}).\]
is even.
Hence if $f$ is cm, the parity of
the complete Bell polynomial in equation (\ref{eq:expf})
is $(-1)^n$ and so $\exp(f)$ is cm. 

The archetypal example of a cm function is $f(x)=\rme^{-x}$.
Other elementary examples of cm functions include:
\begin{enumerate}
\item $f(t)=t^{-\delta}$ ($t>0$) is cm if and only if $\delta\ge0$.
\item $f(t)=\ln(1+t^{-1})$ is cm.
\end{enumerate}
These are proven through
\begin{subequations}\begin{gather}
\frac{\rmd^nx^{-\delta}}{\rmd x^n}=(-\delta)_n^-x^{-\delta-n}
=(-1)^n\frac{(\delta)_n^+}{x^{n+\delta}},\\
\frac{\rmd^{n+1}\ln(1+x^{-1})}{\rmd x^{n+1}}
=(-1)^{n+1}n!\,\biggl[\frac1{x^{n+1}}-\frac1{(1+x)^{n+1}}\biggr].
\end{gather}\end{subequations}
Following this and Lemma \ref{lem:cm} are
\begin{corollary}\label{cor:cmp}
Let $g(t)$ be cm, then both
$t^{-\delta}g(t)$ with $\delta\ge0$ and
$g(t^p)$ with $0<p\le1$ are cm.
\smallskip\\{\it proof.}
The first is obvious thanks to Lemma \ref{lem:cm}-{\it5}.
The last follows Lemma \ref{lem:cm}-{\it6} with $F(t)=t^p$ since
$F'=pt^{p-1}$ for $0<p\le1$ is cm. {\sc q.e.d.}
\end{corollary}
\begin{corollary}\label{cor:qcm}
For $0<p\le1$ and $a,b\ge0$, these are cm:
\[f(t)=t^{-a}(1+t^p)^{-b}\,;\qquad f(t)=t^{-a}(1+t^{-p})^b.\]
{\it proof.}
Let $F(t)=c+t^p$. Then $F'=pt^{p-1}$ is cm for $0<p\le 1$.
Hence first $(g\circ F)(t)=(1+t^p)^{-b}$ with $c=1$ and $g(w)=w^{-b}$
for $0<p\le 1$ and $b\ge0$ is cm.
Next, with $c=0$ and $g(w)=b\ln(1+w^{-1})$,
we find that $(g\circ F)(t)=b\ln(1+t^{-p})$ is cm
for $0<p\le1$ and $b\ge0$,
and so is $(1+t^{-p})^b=\exp[b\ln(1+t^{-p})]$.
The final conclusion follows Corollary \ref{cor:cmp}.
{\sc q.e.d.}
\end{corollary}

The fundamental result characterizing cm functions \citep{HBW,Wi41}
is due to S. Bernstein\footnote{{\cyr\!\!\!\!\!\!Serg\'e\u{i}
Nat\'anovich Bernsht\'e\u{i}n} (1880-1968)},
\begin{theorem}[Hausdorff--\emph{Bernstein}--Widder]\label{th:HBW}\hfill\\
A smooth function $f(x)$ of $x>0$ is completely monotonic
if and only if $f(x)=\int_0^\infty\rme^{-xt}\,\rmd\mu(t)$
where $\mu(t)$ is the Borel measure on $[0,\infty)$,
that is, there exists a non-negative distribution
$\phi(t)\ge0$ of $t>0$ such that equation (\ref{eq:lap}) holds.
\end{theorem}
The `if'-part is elementary since
\[f^{(n)}(x)=(-1)^n\int_0^\infty\!\rmd t\,t^n\rme^{-xt}\phi(t)
=(-1)^n\laplace tx[t^n\phi(t)].\]
Although the complete proof of the `only if'-part
is beyond our scope, the partial proof follows the Post--Widder formula.
That is, if the inverse Laplace transform
$\phi(t)=\mathcal L^{-1}_{x\rightarrow t}[f(x)]$
of a cm function $f(x)$ is well-defined,
then equation (\ref{eq:pinv}), provided that it
converges, indicates that $\phi(t)$
must be non-negative in the positive real domain.

\subsection{Miscellaneous}

We note an additional auxiliary relation, which will be used
throughout this paper: that is, for any non-negative integer $n$ and
arbitrary differentiable function $f(x)$,
\begin{equation}\label{eq:difn}
\biggl(x^2\!\frac\rmd{\rmd x}\biggr)^n(xf)
=x^{n+1}\frac{\rmd^n(x^nf)}{\rmd x^n},
\end{equation}
which may be proven via the induction on $n$ \citep[see][theorem A3]{An11}.
In fact this is also equivalent to a lemma
\begin{equation}\label{eq:lem}
x^nf_{(n+1)}(x)=\frac\rmd{\rmd x}\bigl[x^{n+1}f_{(n)}(x)\bigr]
\end{equation}
where
\[f_{(n)}(x)\equiv\frac{\rmd^n[x^nf(x)]}{\rmd x^n}.\]
This lemma may be proven directly via
\begin{displaymath}\begin{split}
f_{(n+1)}(x)
&=\frac{\rmd^n}{\rmd x^n}\biggl[\frac{\rmd(x\cdot x^nf)}{\rmd x}\biggr]
=\frac{\rmd^n}{\rmd x^n}\biggl[x^nf+x\frac{\rmd(x^nf)}{\rmd x}\biggr]
\\&=\frac{\rmd^n(x^nf)}{\rmd x^n}
+\sum\nolimits_{k=0}^n\binom nk\,\frac{\rmd^kx}{\rmd x^k}
\frac{\rmd^{n-k}}{\rmd x^{n-k}}\biggl[\frac{\rmd(x^nf)}{\rmd x}\biggr]
\\&=\biggl(1+n+x\frac\rmd{\rmd x}\biggr)\,\frac{\rmd^n[x^nf(x)]}{\rmd x^n}
=\frac1{x^n}\frac\rmd{\rmd x}\bigl[x^{n+1}f_{(n)}(x)\bigr]
\end{split}\end{displaymath}
where we also used that $\rmd^kx/\rmd x^k=0$ if $k\ge2$ and
the Leibniz rule (eq.~\ref{eq:leibniz}).
The theorem in equation (\ref{eq:difn}) implying the lemma
in equation (\ref{eq:lem}) has been shown in \citet[corollary A4]{An11}
whereas the opposite implication may be deduced because
the induction step for the proof of equation (\ref{eq:difn}) follows
equation (\ref{eq:lem}) as
\[\frac{\rmd^{n+1}(x^{n+1}f)}{\rmd x^{n+1}}
=\frac1{x^n}
\frac\rmd{\rmd x}\biggl[\Bigl(x^2\!\frac\rmd{\rmd x}\Bigr)^n(xf)\biggr]
=\frac1{x^{n+2}}\biggl(x^2\!\frac\rmd{\rmd x}\biggr)^{n+1}(xf).\]

Fractional calculus also generalizes
the lemma in equation (\ref{eq:lem}) generalizes.
In particular, for a non-negative
integer $n$ and $0\le\delta<1$,
\begin{subequations}\label{eq:dndxnd}
\begin{gather}\begin{split}
\RLI0x{1-\delta}(x^{n+\delta}f)
&=\frac1{\Gamma(1-\delta)}\!
\int_0^x\!\frac{y^{n+\delta}f(y)\,\rmd y}{(x-y)^\delta}
\\&=\frac{x^{n+1}}{\Gamma(1-\delta)}\!
\int_0^1\!\frac{t^{n+\delta}f(xt)\,\rmd t}{(1-t)^\delta}
\end{split}\\\begin{split}
\frd0x{n+\delta}(x^{n+\delta}f)
&=\frac1{\Gamma(1-\delta)}\!
\int_0^1\!\frac{\rmd t\,t^{n+\delta}}{(1-t)^\delta}
\frac{\rmd^{n+1}[x^{n+1}f(xt)]}{\rmd x^{n+1}}
\\&=\frac1{x^{n+1}\Gamma(1-\delta)}\!
\int_0^x\!\frac{y^{n+\delta}f_{(n+1)}(y)\,\rmd y}{(x-y)^\delta}
\\&=\frac1{x^{n+1}}\,
\RLI0x{1-\delta}\bigl[x^{n+\delta}f_{(n+1)}(x)\bigr]
\end{split}\\
x^{n+\delta}f_{(n+1)}(x)
=\frd0x{1-\delta}
\bigl[x^{n+1}\frd0x{n+\delta}(x^{n+\delta}f)\bigr].
\end{gather}
\end{subequations}
Note that for the $\delta=0$, the last results in
equation (\ref{eq:lem}). The middle for the same case is consistent with
the fundamental theorem of calculus given equation (\ref{eq:lem}) indicating
\begin{equation}\label{cor:des}
x^{n+1}f_{(n)}(x)=\cancel{x^{n+1}f_{(n)}(x)\bigr\rvert_{x=0}}
+\int_0^x\!y^nf_{(n+1)}(y)\,\rmd y
\end{equation}
provided that $f_{(n)}(0)$ is finite.
Equations (\ref{eq:lem}) and (\ref{eq:dndxnd}) imply
\begin{corollary}\label{cor:pdes}
for a non-negative integer $n$,
if $f_{(n+1)}(x)\ge0$ for $x>0$,
then $\frd0x\mu(x^\mu f)\ge0$ for $x>0$ and $n\le\mu\le n+1$.
\end{corollary}
In fact, the successive applications of this with a descending subscript
furthermore suggest
that, if $f_{(n)}(x)\ge0$ for $x>0$ and a non-negative
integer $n$, it follows that $\frd0x\mu(x^\mu f)\ge0$ for $x>0$
and any ${}^\forall\mu\le n$.

Corollary \ref{cor:pdes} with an integer $\mu$ may be generalized alternatively, namely,
\begin{theorem}\label{th:cmdes}
for a non-negative integer $n$,
if $x^af_{(n+1)}(x)$ is cm,
then $x^af_{(n)}(x)$ is also cm.
\smallskip\\{\it proof.}
If $x^af_{(n+1)}$ is cm, then by the Bernstein theorem,
there exists a non-negative function $h(u)\ge0$ of $u>0$ such that
\[x^af_{(n+1)}(x)=\int_0^\infty\!\rmd u\,\rme^{-xu}h(u).\]
The complete monotonicity of $x^af_{(n)}$ can then be shown
directly using equation (\ref{cor:des}), which indicates that
\begin{gather*}
x^af_{(n)}=x^{a-n-1}\!\int_0^x\!\rmd y\,y^nf_{(n+1)}(y)
=\int_0^1\!\rmd t\,t^{n-a}\!\int_0^\infty\!\rmd u\,\rme^{-xtu}h(u),
\nonumber\\\frac{\rmd^k[x^af_{(n)}]}{\rmd x^k}
=(-1)^k\int_0^1\!\rmd t\,t^{n+k-a}\!
\int_0^\infty\!\rmd u\,\rme^{-xtu}u^kh(u)
\qquad\square.
\end{gather*}
\end{theorem}

Finally, we also note
\begin{lemma}\label{th:p0}
for a non-negative integer $n$,
if $f^{(n+1)}(a)$ is finite and
$f^{(0)}(a)=\dotsb=f^{(k)}(a)=0$,
then $\frd ax{n+\delta}f(a)=0$ for $0\le\delta<1$.
\smallskip\\{\it proof.}
Here we assume $a=0$, but the similar argument holds for any finite ``$a$''
accompanied by a simple translation. First,
\begin{subequations}\begin{gather}
\RLI 0x{1-\delta}f
=\frac{x^{1-\delta}}{\Gamma(1-\delta)}\!
\int_0^1\!\frac{f(xt)\,\rmd t}{(1-t)^\delta}
\,;\\\label{eq:pd0}
\frd0x{n+\delta}f
=\frac1{\Gamma(1-\delta)}\!\int_0^1\!
\frac{\rmd^{n+1}[y^{1-\delta}f(y)]}{\rmd y^{n+1}}
\biggr\rvert_{y=xt}\frac{t^{n+\delta}\,\rmd t}{(1-t)^\delta}.
\end{gather}
Here the latter follows the former because
\[\frac{\rmd^{n+1}[x^{1-\delta}f(xt)]}{\rmd x^{n+1}}
=t^{n+\delta}\frac{\rmd^{n+1}[y^{1-\delta}f(y)]}{\rmd y^{n+1}}
\biggr\rvert_{y=xt}.\]
Finally, given the Leibniz rule,
\begin{multline*}
\frac{\rmd^{n+1}[y^{1-\delta}f(y)]}{\rmd y^{n+1}}
=y^{1-\delta}f^{(n+1)}(y)
\\+(1-\delta)\sum_{k=0}^n(-1)^{n-k}\binom{n+1}k\,(\delta)_{n-k}^+
\frac{f^{(k)}(y)}{y^{n+\delta-k}},
\end{multline*}\end{subequations}
which identically vanishes for $y=0$ if
the condition part of Lemma \ref{th:p0} with $a=0$ holds. Here
the conclusion follows as the integrand of equation (\ref{eq:pd0})
with $x=0$ is also zero. {\sc q.e.d.}
\end{lemma}

\section{Fractional calculus on the augmented density}
\label{sec:cal}

\citet{An11a} has shown that the Abel transformation of the
augmented moment function 
of an anisotropic spherical system results in a similar integral
transformation of the df as equation (\ref{eq:dist}) but with
different powers on $\mathcal K$ and $L^2$. This result
generalizes by means of the fractional calculus. The goal
of this section is to establish them
(see eqs.~\ref{eq:dpirn2} and \ref{eq:dripn2})
for any pair of non-negative reals $0\le\mu\le\xi$.

We start by
considering to apply the integral operator of equation (\ref{eq:mint})
to equation (\ref{eq:dist}) on $\Psi$ or $r^2$. In fact, we can
establish more general results. With
\begin{subequations}\label{eq:ints}
\[
\mathscr I_s(\Psi,r^2)\equiv
\iint_\mathcal T\!\rmd\mathcal E\,\rmd L^2\mathcal K^s
G(\mathcal E,L^2)
\]
where the $\Psi$ and $r^2$ dependencies of the integrable function
$G=G(\mathcal E,L^2)$ are only through the two integrals of motion
$\mathcal E$ and $L^2$
(henceforth these trivial arguments of $G$ will be suppressed
for the sake of brevity), the Fubini theorem implies
\begin{align*}
\RLI{\mathcal E_0}\Psi\lambda&\mathscr I_s
=\iint_{\mathcal E\ge\mathcal E_0,L^2\ge0}\!\rmd\mathcal E\,\rmd L^2G\,
\RLI{\mathcal E_0}\Psi\lambda
\bigl[\mathcal K^s\Theta(\mathcal K)\bigr],
\\
\RLI0{r^2}\lambda&\biggl(\frac{\mathscr I_s}{r^{2\lambda+2}}\biggr)
=\iint_{\mathcal E\ge\mathcal E_0,L^2\ge0}\!\rmd\mathcal E\,\rmd L^2G\,
\RLI0{r^2}\lambda
\biggl[\frac{\mathcal K^s\Theta(\mathcal K)}{r^{2\lambda+2}}\biggr],
\\
\RLI0{r^2}\lambda&\Bigl(r^{2s}\!\mathscr I_s\Bigr)
=\iint_{\mathcal E\ge\mathcal E_0,L^2\ge0}\!\rmd\mathcal E\,\rmd L^2G\,
\RLI0{r^2}\lambda
\bigl[r^{2s}\mathcal K^s\Theta(\mathcal K)\bigr].
\end{align*}
Through direct calculations that are basically identical to
that of \citet[appendix A]{An11a} except for
different arguments of the Euler integral for
the beta function, we find that
\begin{gather*}\begin{split}
\RLI{\mathcal E_0}\Psi\lambda&
\bigl[\mathcal K^s\Theta(\mathcal K)\bigr]
\\&=\frac{\Theta(\mathcal K)}{\Gamma(\lambda)}\!
\int_{\mathcal E+\frac{L^2}{2r^2}}^\Psi\!\rmd Q\,(\Psi-Q)^{\lambda-1}
\biggl[2(Q-\mathcal E)-\frac{L^2}{r^2}\biggr]^s
\\&=\frac{\Theta(\mathcal K)}{\Gamma(\lambda)}
\frac{\mathcal K^{s+\lambda}}{2^\lambda}\mathrm B(\lambda,s+1),
\end{split}\\\begin{split}
\RLI0{r^2}\lambda&
\biggl[\frac{\mathcal K^s\Theta(\mathcal K)}{r^{2\lambda+2}}\biggr]
\\&=\frac{\Theta(\mathcal K)}{\Gamma(\lambda)}\!
\int_{\frac{L^2}{2(Q-\mathcal E)}}^{r^2}\!\rmd R^2
\frac{(r^2-R^2)^{\lambda-1}}{R^{2\lambda+2}}
\biggl[2(\Psi-\mathcal E)-\frac{L^2}{R^2}\biggr]^s
\\&=\frac{\Theta(\mathcal K)}{\Gamma(\lambda)}
\frac{r^{2\lambda-2}\mathcal K^{s+\lambda}}{L^{2\lambda}}
\mathrm B(\lambda,s+1),
\end{split}\\\begin{split}
\RLI0{r^2}\lambda&
\bigl[r^{2s}\mathcal K^s\Theta(\mathcal K)\bigr]
\\&=\frac{\Theta(\mathcal K)}{\Gamma(\lambda)}\!
\int_{\frac{L^2}{2(Q-\mathcal E)}}^{r^2}\!\rmd R^2
R^{2s}(r^2-R^2)^{\lambda-1}
\biggl[2(\Psi-\mathcal E)-\frac{L^2}{R^2}\biggr]^s
\\&=\frac{\Theta(\mathcal K)}{\Gamma(\lambda)}
\frac{\mathcal K^{s+\lambda}}{2^\lambda(\Psi-\mathcal E)^\lambda}
\mathrm B(\lambda,s+1).
\end{split}\end{gather*}
Hence, we have established that
\begin{align}
\RLI{\mathcal E_0}\Psi\lambda&\mathscr I_s
=\frac{\Gamma(s+1)}{2^\lambda\Gamma(s+\lambda+1)}\!
\iint_\mathcal T\!\rmd\mathcal E\,\rmd L^2
\mathcal K^{s+\lambda}G,
\\
\RLI0{r^2}\lambda&\biggl(\frac{\mathscr I_s}{r^{2\lambda+2}}\biggr)
=\frac{r^{2\lambda-2}\Gamma(s+1)}{\Gamma(s+\lambda+1)}\!
\iint_\mathcal T\!\rmd\mathcal E\,\rmd L^2
\frac{\mathcal K^{s+\lambda}G}{L^{2\lambda}},
\\\label{eq:ints0}
\RLI0{r^2}\lambda&\Bigl(r^{2s}\!\mathscr I_s\Bigr)
=\frac{r^{2(s+\lambda)}\Gamma(s+1)}{2^\lambda\Gamma(s+\lambda+1)}\!\!
\iint_\mathcal T\!\!
\frac{\mathcal K^{s+\lambda}G\rmd\mathcal E\,\rmd L^2}
{(\Psi-\mathcal E)^\lambda},
\end{align}
\end{subequations}
which are valid for any $s>-1$ and $\lambda\ge0$, provided that
all integrals on the right-hand sides converge.

We next find differentiations of the integral transform
$\mathscr I_s$, namely
(here $X\equiv\Psi$ or $r^2$)
\begin{equation}\label{eq:difs}
\frac{\partial\mathscr I_s}{\partial X}
=\begin{cases}{\displaystyle
s\!\iint_\mathcal T\!\rmd\mathcal E\,\rmd L^2\mathcal K^{s-1}
\frac{\partial\mathcal K}{\partial X}G
}&(s>0)\smallskip\\
\frac12\!{\displaystyle\int_0^{\bar L^2}\!\rmd L^2
\frac{\partial\mathcal K}{\partial X}\biggr\rvert_{\mathcal K=0}
G\Bigl(\Psi-\frac{L^2}{2r^2},L^2\Bigr)
}&(s=0)\end{cases}.
\end{equation}
The $\frac12$-factor for the $s=0$ case is due to
\[\deltaup(\mathcal K)=
\tfrac12\deltaup\Bigl[\Psi-\frac{L^2}{2r^2}-\mathcal E\Bigr]\]
where $\deltaup(x)=\Theta^\prime(x)$ is the Dirac delta.
In addition, \[\bar L^2=\begin{cases}
2r^2\Psi&\text{if $\mathcal E_0=0$}\\
\infty&\text{if $\mathcal E_0=-\infty$}\end{cases}.\]
Given that \[\frac{\partial\mathcal K}{\partial\Psi}=2\,;\qquad
\frac{\partial\mathcal K}{\partial r^2}=\frac{L^2}{r^4},\]
equation (\ref{eq:difs}) suggests that
for an integer $n\ge0$ and $s>-1$,
\begin{subequations}\label{eq:difsn}
\begin{gather}
\frac{\partial^n\!\mathscr I_s}{\partial\Psi^n}
=\begin{cases}{\displaystyle
2^n(s)_n^-\!\iint_\mathcal T\!\rmd\mathcal E\,\rmd L^2\mathcal K^{s-n}G
}&(n<s+1)\\{\displaystyle
2^s\Gamma(1+s)\!\!\int_0^{\bar L^2}\!\!\rmd L^2
G\Bigl(\Psi-\frac{L^2}{2r^2},L^2\Bigr)
}&(n=s+1)\end{cases},
\\
\begin{split}\label{eq:difsnr}
\biggl(r^4\!&\frac\partial{\partial r^2}\biggr)^n\mathscr I_s
\\&=\begin{cases}{\displaystyle
(s)_n^-\!\iint_\mathcal T\!\rmd\mathcal E\,\rmd L^2\mathcal K^{s-n}L^{2n}G
}&(n<s+1)\\{\displaystyle
\frac{\Gamma(1+s)}2\!\!\int_0^{\bar L^2}\!\!\rmd L^2
L^{2s+2}G\Bigl(\Psi-\frac{L^2}{2r^2},L^2\Bigr)
}&(n=s+1)\end{cases}.
\end{split}\end{gather}
\end{subequations}

Equations (\ref{eq:ints}), (\ref{eq:difsn})
and $\Nu=m_{0,0}$ expressed as an integral transformation of the df
as in equation (\ref{eq:dist}) result in
\begin{align}\label{eq:dpirn}
&\frac{\partial^n}{\partial\Psi^n}\biggl[
\RLI0{r^2}{\xi-\frac12}\Bigl(\frac{\Nu}{r^{2\xi-1}}\Bigr)\biggr]
\\\nonumber&\quad=\begin{cases}{\displaystyle
\frac{2^{n+1}\pi^\frac32r^{2\xi-3}}{\Gamma(\xi-n)}\!
\iint_\mathcal T\!\rmd\mathcal E\,\rmd L^2
\frac{\mathcal K^{\xi-n-1}}{L^{2\xi-1}}
\mathcal F(\mathcal E,L^2)
}&(n<\xi)\smallskip\\{\displaystyle
2^\xi\pi^\frac32r^{2\xi-3}\!
\int_0^{\bar L^2}\!\frac{\rmd L^2}{L^{2\xi-1}}
\mathcal F\Bigl(\Psi-\frac{L^2}{2r^2},L^2\Bigr)
}&(n=\xi)\end{cases},
\\\label{eq:dripn}
&\biggl(r^4\!\frac\partial{\partial r^2}\biggr)^n
\Bigl(r^2\RLI{\mathcal E_0}\Psi{\xi-\frac12}\Nu\Bigr)
\\\nonumber&\quad=\begin{cases}{\displaystyle
\frac{2^{\frac32-\xi}\pi^\frac32}{\Gamma(\xi-n)}\!
\iint_\mathcal T\!\rmd\mathcal E\,\rmd L^2
\mathcal K^{\xi-n-1}L^{2n}\mathcal F(\mathcal E,L^2)
}&(n<\xi)\smallskip\\{\displaystyle
2^{\frac12-\xi}\pi^\frac32\!\int_0^{\bar L^2}\!\rmd L^2
L^{2\xi}\mathcal F\Bigl(\Psi-\frac{L^2}{2r^2},L^2\Bigr)
}&(n=\xi)\end{cases}.
\end{align}
where $n$ is again a non-negative integer and $\xi\ge\frac12$.
Both equations further generalize from an integer $n$ to a real $\mu\le\xi$
using fractional order derivatives, and it can also be
shown that they are in fact valid for $\xi\ge0$ if the extended
definition in equation (\ref{eq:mintn}) is adopted.

In particular, to generalize equation (\ref{eq:dpirn}),
we first find that
\begin{equation}\label{eq:int2}
\RLI{\mathcal E_0}\Psi\lambda\biggl[
\RLI0{r^2}{\xi-\frac12}\Bigl(\frac{\Nu}{r^{2\xi-1}}\Bigr)\biggr]
=\frac{2\pi^\frac32r^{2\xi-3}}{2^\lambda\Gamma(\xi+\lambda)}\!
\iint\limits_\mathcal T\!\rmd\mathcal E\,\rmd L^2\!
\frac{\mathcal K^{\lambda+\xi-1}}{L^{2\xi-1}}
\mathcal F(\mathcal E,L^2)
\end{equation}
for $\xi\ge\frac12$ and $\lambda\ge0$,
which follows equation (\ref{eq:ints}).
The generalization of equation (\ref{eq:dpirn}) is arrived
by applying equation (\ref{eq:difsn}), that is,
for any reals $0\le\mu\le\xi$ and $\xi\ge\frac12$
(the latter restriction that $\xi\ge\frac12$ will be dropped later
in this section),
\begin{align}\label{eq:dpirn2}
&\frd{\mathcal E_0}\Psi\mu\biggl[
\RLI0{r^2}{\xi-\frac12}_{r^2}\Bigl(\frac{\Nu}{r^{2\xi-1}}\Bigr)\biggr]
\\\nonumber&\quad=\begin{cases}{\displaystyle
\frac{2^{\mu+1}\pi^\frac32r^{2\xi-3}}{\Gamma(\xi-\mu)}\!
\iint_\mathcal T\!\rmd\mathcal E\,\rmd L^2
\frac{\mathcal K^{\xi-\mu-1}}{L^{2\xi-1}}
\mathcal F(\mathcal E,L^2)
}&(\mu<\xi)\smallskip\\{\displaystyle
2^\xi\pi^\frac32r^{2\xi-3}\!
\int_0^{\bar L^2}\!\frac{\rmd L^2}{L^{2\xi-1}}
\mathcal F\Bigl(\Psi-\frac{L^2}{2r^2},\,L^2\Bigr)
}&(\mu=\xi)\end{cases},
\end{align}
provided that the integrals converge.
Equation (\ref{eq:dpirn2}) for $\xi=\frac12$ now reduces to
\begin{equation}\label{eq:dmn}
\!r^2\,\frd{\mathcal E_0}\Psi\mu\Nu
=\begin{cases}{\displaystyle
\frac{2^{1+\mu}\pi^\frac32}{\Gamma(\frac12-\mu)}\!\!
\iint_\mathcal T\!\!\rmd\mathcal E\,\rmd L^2
\frac{\mathcal F(\mathcal E,L^2)}{\mathcal K^{\mu+\frac12}}
}&(\mu<\frac12)\smallskip\\{\displaystyle
\sqrt2\pi^{\frac32}\!\!\int_0^{\bar L^2}\!\!\rmd L^2
\mathcal F\Bigl(\Psi-\frac{L^2}{2r^2},L^2\Bigr)
}&(\mu=\frac12)\end{cases}.
\end{equation}
Here setting $\mu=\frac12-\xi$ results in
equation (\ref{eq:dripn}) with $n=0$ given that
$\RLI{\mathcal E_0}\Psi{\xi-\frac12}\Nu
=\frd{\mathcal E_0}\Psi{\frac12-\xi}\Nu$.
It is inferred that equation (\ref{eq:dripn}) is in fact valid
for $\xi\ge0$ (n.b., $0\le n\le\xi$ and so
if $0\le\xi\le\frac12$, then $n=0$).

A similar generalization of equation (\ref{eq:dripn})
from an integer $n$ to a real $\mu$ (cf., eq.~\ref{eq:difn}) and
the extension of equation (\ref{eq:dpirn2}) to $\xi\ge0$ are
possible although demonstrating them through direct calculations
is comparatively nontrivial. Instead, we derive the generalization
of equation (\ref{eq:dripn}) following an indirect route.
Let us first consider combining equation (\ref{eq:ints0})
with $G=\mathcal F$, $\mu=s+1>0$ and $\lambda=1-\delta$
where $\delta=\mu-\flr\mu$, and 
equation (\ref{eq:dripn}) with $n=0$ and $\xi=\mu>0$, which results in
\[
\RLI0{r^2}{\!1-\delta}\Bigl(
r^{2\mu}\,\RLI{\mathcal E_0}\Psi{\!\mu-\frac12}\Nu\Bigr)
=\frac{2^{\frac12-\flr\mu}\pi^\frac32r^{2\flr\mu}}
{\Gamma(1+\flr\mu)}\!\!
\iint\limits_\mathcal T\!\!\rmd\mathcal E\,\rmd L^2
\frac{\mathcal K^{\flr\mu}\mathcal F(\mathcal E,L^2)}
{(\Psi-\mathcal E)^{1-\delta}}
\]
for $\mu>0$ and $0<\delta<1$.
Next equation (\ref{eq:difsnr}) indicates that
\begin{multline*}
\biggl(r^4\!\frac\partial{\partial r^2}\biggr)^{n+1}
\biggl[\frac1{r^{2\flr\mu}}\RLI0{r^2}{\!1-\delta}\Bigl(
r^{2\mu}\,\RLI{\mathcal E_0}\Psi{\mu-\frac12}\Nu\Bigr)\biggr]
\\=\frac{\pi^\frac32r^{2-2\delta}}{2^{\mu-\frac12}}\!
\int_0^{\bar L^2}\!\rmd L^2L^{2\mu}
\mathcal F\Bigl(\Psi-\frac{L^2}{2r^2},L^2\Bigr).
\end{multline*}
for a non-negative integer $n=\flr\mu$. However,
\begin{multline*}
\biggl(r^4\!\frac\partial{\partial r^2}\biggr)^{n+1}
\biggl[\frac1{r^{2\flr\mu}}\RLI0{r^2}{\!1-\delta}\Bigl(
r^{2\mu}\,\RLI{\mathcal E_0}\Psi{\mu-\frac12}\Nu\Bigr)\biggr]
\\=r^{2\flr\mu+4}\,
\biggl(\frac\partial{\partial r^2}\biggr)^{\flr\mu+1}
\RLI0{r^2}{\!1-\delta}\Bigl(
r^{2\mu}\,\RLI{\mathcal E_0}\Psi{\mu-\frac12}\Nu\Bigr)
\end{multline*}
thanks to equation (\ref{eq:difn}),
and consequently, we find that
\begin{subequations}
\begin{equation}\label{eq:lmom0}
\frd0{r^2}\mu\Bigl(
r^{2\mu}\,\RLI{\mathcal E_0}\Psi{\mu-\frac12}\Nu\Bigr)
=\frac{\pi^\frac32}{2^{\mu-\frac12}r^{2\mu+2}}\!
\int_0^{\bar L^2}\!\rmd L^2L^{2\mu}
\mathcal F\Bigl(\Psi-\frac{L^2}{2r^2},L^2\Bigr).
\end{equation}
This is also consistent with the case $n=\xi$ of
equation (\ref{eq:dripn}), again thanks to equation (\ref{eq:difn}).
That is to say, equation (\ref{eq:lmom0}) is actually valid
for any $\mu\ge0$ including integer values.

Finally, consider applying the integral operator in
equation (\ref{eq:mint}) on $\Psi$ to equation (\ref{eq:lmom0}), as in
\[
\RLI{\mathcal E_0}\Psi{\xi-\mu}
\biggl[\frd0{r^2}\mu_{r^2}\Bigl(
r^{2\mu}\,\RLI{\mathcal E_0}\Psi{\mu-\frac12}\Nu\Bigr)\biggr]
=\frd0{r^2}\mu_{r^2}\Bigl(
r^{2\mu}\,\RLI{\mathcal E_0}\Psi{\xi-\frac12}\Nu\Bigr)
\]
where $\xi\ge\mu$.
The actual calculations is aided by an alternative expression for
the right-hand side of equation (\ref{eq:lmom0})
\begin{equation}
\frd0{r^2}\mu\Bigl(
r^{2\mu}\,\RLI{\mathcal E_0}\Psi{\mu-\frac12}\Nu\Bigr)
=(2\pi)^\frac32\!
\int_{\mathcal E_0}^\Psi\!\rmd\mathcal E\,(\Psi-\mathcal E)^\mu
\mathcal F\bigl[\mathcal E,2r^2(\Psi-\mathcal E)\bigr].
\end{equation}
It then follows that for $0\le\mu<\xi$
\begin{multline}\label{eq:lmom}
\frac{\Gamma(\xi-\mu)}{(2\pi)^\frac32}\frd0{r^2}\mu_{r^2}\Bigl(
r^{2\mu}\,\RLI{\mathcal E_0}\Psi{\xi-\frac12}\Nu\Bigr)
\\\quad
=\int_{\mathcal E_0}^\Psi\!\rmd Q\,(\Psi-Q)^{\xi-\mu-1}\!
\int_{\mathcal E_0}^Q\!\rmd\mathcal E\,
(Q-\mathcal E)^\mu\mathcal F\bigl[\mathcal E,\,2r^2(Q-\mathcal E)\bigr]
\\\quad
=\int_{\mathcal E_0}^\Psi\!\rmd\mathcal E\!
\int_\mathcal E^\Psi\!\rmd Q\,
(\Psi-Q)^{\xi-\mu-1}(Q-\mathcal E)^\mu
\mathcal F\bigl[\mathcal E,\,2r^2(Q-\mathcal E)\bigr]
\\\quad
=\frac1{(2r^2)^{\mu+1}}\!\!
\int_{\mathcal E_0}^\Psi\!\!\rmd\mathcal E\!
\int_0^{2r^2(\Psi-\mathcal E)}\!\rmd L^2
\Bigl(\Psi-\mathcal E-\frac{L^2}{2r^2}\Bigr)^{\xi-\mu-1}\!
L^{2\mu}\mathcal F(\mathcal E,L^2)
\\=\frac1{2^\xi r^{2\mu+2}}\!
\iint\limits_\mathcal T\!\rmd\mathcal E\,\rmd L^2
\mathcal K^{\xi-\mu-1}L^{2\mu}\mathcal F(\mathcal E,L^2).
\end{multline}
\end{subequations}
Equations (\ref{eq:lmom0}) and (\ref{eq:lmom}) together, that is,
\begin{align}\label{eq:dripn2}
&\frd0{r^2}\mu_{r^2}\Bigl(
r^{2\mu}\,\RLI{\mathcal E_0}\Psi{\xi-\frac12}\Nu\Bigr)
\\\nonumber&\quad=\begin{cases}{\displaystyle
\frac{2^{\frac32-\xi}\pi^\frac32}{r^{2\mu+2}\Gamma(\xi-\mu)}\!
\iint_\mathcal T\!\rmd\mathcal E\,\rmd L^2
\mathcal K^{\xi-\mu-1}L^{2\mu}\mathcal F(\mathcal E,L^2)
}&(\xi>\mu)\smallskip\\{\displaystyle
\frac{\pi^\frac32}{2^{\mu-\frac12}r^{2\mu+2}}\!
\int_0^{\bar L^2}\!\rmd L^2L^{2\mu}
\mathcal F\Bigl(\Psi-\frac{L^2}{2r^2},\,L^2\Bigr)
}&(\xi=\mu)\end{cases}
\end{align}
constitute the generalization of equation (\ref{eq:dripn})
from an integer $n$ to a real $\mu$, which is valid
for any pair of $\mu$ and $\xi$ with $0\le\mu\le\xi$.
For $0\le\mu\le\xi\le\frac12$, the indices transform
$(\mu,\xi)\rightarrow(\frac12-\xi,\frac12-\mu)$ sends
equation (\ref{eq:dripn2}) to (\ref{eq:dpirn2}) given
equation (\ref{eq:mintn}).
Equations (\ref{eq:dpirn2}) and (\ref{eq:dripn2}) thus are
both valid for any real pair $\mu$ and $\xi$ with $0\le\mu\le\xi$.

In fact, both results and also equation \ref{eq:int2} are
different manifestations of the same result, that is to say,
\begin{align}\label{eq:last}
&\frac{r^2}{\sqrt2\pi^\frac32}
\RLI{\mathcal E_0}\Psi\lambda\,
\RLI0{r^2}\xi\Bigl(\frac{\Nu}{r^{2\xi}}\Bigr)
\\\nonumber&\quad=\begin{cases}{\displaystyle
\frac{2^{\frac12-\lambda}r^{2\xi}}{\Gamma(\lambda+\xi+\frac12)}\!\!
\iint_\mathcal T\!\!\rmd\mathcal E\,\rmd L^2
\frac{\mathcal K^{\lambda+\xi-\frac12}\mathcal F(\mathcal E,L^2)}{L^{2\xi}}
}&(\lambda+\xi>-\frac12)\\{\displaystyle
(2r^2)^\xi\!\int_0^{\bar L^2}\!\frac{\rmd L^2}{L^{2\xi}}
\mathcal F\Bigl(\Psi-\frac{L^2}{2r^2},L^2\Bigr)
}&(\lambda+\xi=-\frac12)\end{cases}.
\end{align}
which are valid for any real pair $(\lambda,\xi)$ such that
$\lambda+\xi+\frac12\ge0$.

\newpage
\section{Moment sequences and augmented densities}
\label{sec:moms}

Consider the moment sequence of the df in $(\mathcal E,L^2)$ space
restricted along $\mathcal K=0$, given as in
\begin{subequations}
\begin{align}
F_\mu&(\Psi,r^2)\equiv\frac{(2\pi)^\frac32}{(2r^2)^{\mu+1}}\!
\int_0^{\bar L^2}\!\rmd L^2
L^{2\mu}\mathcal F\Bigl(\Psi-\frac{L^2}{2r^2},L^2\Bigr)
\label{eq:moms}\\\nonumber&
=\begin{cases}{\displaystyle
\Psi^{\mu+1}\!\int_0^1\!\rmd y\,y^\mu\mathscr F(y\Psi;\Psi,r^2)
}&(\mathcal E_0=0,\,\bar L^2=2r^2\Psi)
\\{\displaystyle
\int_0^\infty\!\rmd Y\,Y^\mu\mathscr F(Y;\Psi,r^2)
}&(\mathcal E_0=-\infty,\,\bar L^2=\infty)
\end{cases},
\end{align}
where
\begin{equation}
\mathscr F(Y;\Psi,r^2)\equiv
(2\pi)^\frac32\mathcal F(\Psi-Y,2r^2Y).
\end{equation}
\end{subequations}
Then equations (\ref{eq:dpirn2}) and (\ref{eq:dripn2}) indicate that
\begin{equation}
\label{eq:msad}
F_\mu=\begin{cases}
\RLI{\mathcal E_0}\Psi{\mu-\frac12}
\frd0{r^2}\mu\bigl(r^{2\mu}\Nu\bigr)
&(\mu\ge\frac12)\\
\frd{\mathcal E_0}\Psi{\frac12-\mu}
\frd0{r^2}\mu\bigr(r^{2\mu}\Nu\bigr)
&(0\le\mu\le\frac12)\\
\frd{\mathcal E_0}\Psi{\xi+\frac12}
\RLI0{r^2}\xi
\Bigl(\dfrac{\Nu}{r^{2\xi}}\Bigr)
&(\xi=-\mu\ge0)\end{cases}.
\end{equation}
In particular, if $\mu$ is a positive integer,
this results in
\begin{align*}
F_0&=\frac1{\sqrt\pi}\frac\partial{\partial\Psi}\!
\int_{\mathcal E_0}^\Psi\!
\frac{\Nu(Q,r^2)\,\rmd Q}{\sqrt{\Psi-Q}}
\\F_n&=\frac1{\bigl(\tfrac12\bigr)_{n-1}^+\!\sqrt\pi}
\int_{\mathcal E_0}^\Psi\!\rmd Q\,
(\Psi-Q)^{n-\frac32}
\biggl(\frac\partial{\partial r^2}\biggr)^n\bigl[r^{2n}\Nu(Q,r^2)\bigr],
\end{align*}
where $n=1,2,\dotsc$.
That is to say, a set of fractional calculus chains of the AD
directly determine the entire moment sequences along a fixed
sectional line in $(\mathcal E,L^2)$ space.
In other words, the AD is similar
to the moment generating function (or the characteristic function)
for the df as a probability density.
With varying $(\Psi,r^2)$, the $\mathcal K=0$ lines
eventually sweep the whole accessible $(\mathcal E,L^2)$ space,
and thus $\Nu(\Psi,r^2)$ in principle uniquely determine
the two-integral df, $f(\mathcal E,L^2)$.
A few explicit inversion algorithms
from $\Nu(\Psi,r^2)$ to $f(\mathcal E,L^2)$ are already
available in the literature utilizing either the known inverse of
named integral transforms \cite[see e.g.,][]{Ly62,De86,BvH07} or
complex contour integrals \cite[see e.g.,][]{HQ93,An11a}.
Since the definition of the AD in equation (\ref{eq:ad})
provides the explicit formula from $f(\mathcal E,L^2)$ to
$\Nu(\Psi,r^2)$, the knowledge of $\Nu(\Psi,r^2)$ is therefore
mathematically equivalent to knowing $f(\mathcal E,L^2)$.
Once the potential $\Psi=\Psi(r)$
is specified, the specification of the AD thus completely
determine a unique spherical dynamic system in equilibrium.
Although this approach to the df $f(\mathcal E,L^2)$ through
the AD $\Nu(\Psi,r^2)$ is advantageous as the observables
constrain the AD more directly than the df, this procedure suffers
a significant drawback in that the df recovered as such is indeed
physical, that is, non-negative everywhere in the all accessible
subvolume of the phase-space -- the ``\emph{phase-space consistency}'',
which is the subject of the reminder of this paper following the
current chapter.

Next, we consider what information on the physical properties
of the system is sufficient to specify a unique AD. First,
we find from equation (\ref{eq:dripn2}) that
the (augmented) velocity moments of the even orders are related to
the AD as in
\begin{equation}\label{eq:vmts}\begin{split}
m_{k,n}(\Psi,r^2)&
=\frac{2^{k+n}\Gamma(k+\frac12)}{\sqrt\pi r^{2n+2}}
\biggl(r^4\!\frac\partial{\partial r^2}\biggr)^n
\bigl(r^2\,\RLI{\mathcal E_0}\Psi{n+k}\Nu\bigr)
\\&=2^{k+n}\bigl(\tfrac12\bigr)_k^+
\RLI{\mathcal E_0}\Psi{k+n}
\bigl[\frd0{r^2}n(r^{2n}\Nu)\bigr].
\end{split}\end{equation}
Here note that $(\frac12)_k^+=\Gamma(k+\frac12)/\sqrt\pi$.
This is basically equation (13) of \citet{DM92}
-- see also equation (8) of \citet{BvH07},
equation (A2) of \citet{vHBD09}, equation (5c) of \citet{An11} and so on.
Equation (\ref{eq:vmts}) indicates that, given potential $\Psi(r)$,
specifying the AD completely fixes every
(in principle observable) non-vanishing velocity moment such that
\[\overline{v_r^{2k}v_\mathrm t^{2n}}=
\frac{m_{k,n}[\Psi(r),r^2]}{\Nu[\Psi(r),r^2]}.\]
Conversely, equation (\ref{eq:vmts}) for $(k,n)=(\mu+1,0)$, that is,
$m_{\mu+1,0}=2^{\mu+1}(\frac12)_{\mu+1}^+
\RLI{\mathcal E_0}\Psi{\mu+1}\Nu$
\emph{at a fixed $r$} reduces to
\begin{subequations}
\begin{align}
M_\mu(r)&\equiv
\frac{\mu!\overline{v_r^{2(\mu+1)}}}
{2^{\mu+1}\bigl(\frac12\bigr)_{\mu+1}^+}
\\\nonumber&=\begin{cases}{\displaystyle
[\Psi(r)]^{\mu+1}\!\int_0^1\!\rmd q\,q^\mu\mathscr P\bigl[q\Psi(r);r\bigr]
}&(\mathcal E_0=0)\\{\displaystyle
\int_0^\infty\!\rmd Q\,Q^\mu\mathscr P(Q;r)
}&(\mathcal E_0=-\infty)
\end{cases},
\end{align}
where
\begin{equation}
\mathscr P(Q;r)\equiv\frac{\Nu[\Psi(r)-Q,r^2]}{\nu(r)}.
\end{equation}
\end{subequations}
In other words, given the knowledges of the local density $\nu(r)$ and
the potential $\Psi(r)$, the infinite set of the radial velocity moments
in every order consists in the moment sequence of
the AD considered as a distribution of $\Psi$
-- over the compact support if $\mathcal E_0=0$ or
the half-open interval $[0,\infty)$ if $\mathcal E_0=-\infty$ --
at fixed $r$. The problem is closely related to the
Hausdorff\footnote{Felix Hausdorff (1868-1942)}
(for $\mathcal E_0=0$) or the Stieltjes\footnote{Thomas Joannes
Stieltjes (1856-1894)} (for $\mathcal E_0=-\infty$) moment problems.
With the infinite sequence of the radial velocity moments
as functions of $r$, the AD can then be
uniquely determined at least formally by such means as
e.g., the Hilbert\footnote{David Hilbert (1862-1943)} basis or
the Laplace and/or Fourier\footnote{Jean Baptiste Joseph Fourier
(1768-1830)} transform (cf.,
the moment generating function and the characteristic function) etc.

The final information required for the full specification of the system
is then the determination of the potential.
Clearly the potential may be determined through the Poisson equation
$\nabla^2\Phi=4\pi G\rho$, which under the spherical symmetry reduces to
\begin{equation}\label{eq:poi}
\frac1{r^2}\frac\rmd{\rmd r}\biggl(r^2\frac{\rmd\Psi}{\rmd r}\biggr)
=-4\pi G\Upsilon\nu.
\end{equation}
Hence if $\Upsilon\equiv\rho(r)/\nu(r)$ is assumed to be constant,
$\Psi(r)$ can be fixed by solving the ordinary differential equation
on $\Psi(r)$ that results from setting $\nu=\Nu(\Psi,r^2)$ in
equation (\ref{eq:poi}). Alternatively, from
equation (\ref{eq:vmts}), we deduce for $k\ge1$ that
\begin{subequations}
\begin{equation}\begin{split}
&\frac{\partial m_{k,n}}{\partial\Psi}
=(2k-1)\,m_{k-1,n}
\,;\\
&\frac{\partial(r^{2n+2}m_{k,n})}{\partial r^2}
=\bigl(k-\tfrac12\bigr)\,r^{2n}m_{k-1,n+1}.
\end{split}\end{equation}
\newpage\noindent%
The total radial derivative of $m_{k,n}$ for $k\ge1$ then
results in
\begin{multline}
\frac{\rmd m_{k,n}}{\rmd r}
=\frac{2m_{k,n}}{r}\biggl[
\frac{\partial\log(r^{2n+2}m_{k,n})}{\partial\log r^2}
-(n+1)\biggr]
+\frac{\rmd\Psi}{\rmd r}\frac{\partial m_{k,n}}{\partial\Psi}
\\=-\frac{2(n+1)m_{k,n}-(2k-1)m_{k-1,n+1}}r
\\+(2k-1)m_{k-1,n}\frac{\rmd\Psi}{\rmd r}.
\end{multline}
\end{subequations}
With $\Psi=\Psi(r)$ and
$m_{k,n}[\Psi(r),r^2]=\nu\overline{v_r^{2k}v_\mathrm t^{2n}}$,
this may be solved for $\rmd\Psi/\rmd r$ if the required velocity
moments as a function of $r$ are known. For the simplest case
$(k,n)=(1,0)$, this reduces to the spherical
(second-order steady-state) Jeans equation,
\begin{equation}\label{eq:je}
\frac1\nu\frac{\rmd(\nu\overline{v_r^2})}{\rmd r}
+\frac{2\overline{v_r^2}-\overline{v_\mathrm t^2}}r
=\frac{\rmd\Psi}{\rmd r},
\end{equation}
that is, the spherically-symmetric hydrostatic equilibrium
equation with an anisotropic velocity dispersion tensor.

\section{Necessary condition for separable augmented densities}
\label{sec:nec}

In the following, we limit our concern to the cases for which
the potential and the radius dependencies of the AD are
multiplicatively separable such that
\begin{equation}\label{eq:sep}
\Nu(\Psi,r^2)=P(\Psi)R(r^2).
\end{equation}
In addition to mathematical expediency, this assumption is also notable
because under the separability assumption in equation (\ref{eq:sep}),
the radius part $R(r^2)$ of the AD alone can uniquely specify
the so-called \citeauthor{Bi80} anisotropy parameter,
\begin{gather}\begin{split}
\beta(r)&=1-\frac{\overline{v_\mathrm t^2}}{2\overline{v_r^2}}
=1-\frac{m_{0,1}[\Psi(r),r^2]}{2m_{1,0}[\Psi(r),r^2]}
\\&=1-\frac1{m_{1,0}}\frac{\partial(r^2m_{1,0})}{\partial r^2}
=-\frac{\partial\log m_{1,0}}{\partial\log r^2}\biggr\rvert_{\Psi(r),r^2}
\end{split}
\intertext{such that (\citealt{De86,QH95,BvH07,An11};
see also \citealt{vdM94} as $R^{-1}$ being the integrating factor
of the Jeans equation, i.e, eq.~\ref{eq:je})}
\beta(r)=-\frac{\rmd\log R(r^2)}{\rmd\log r^2}
\,;\qquad
\frac{R(r^2)}{R(r_0^2)}
=\exp\biggl\lgroup\int_r^{r_0}\!\frac{2\beta(s)}s\rmd s\biggr\rgroup.
\label{eq:beta}\end{gather}
Some applications are found e.g., in \citet{BvH07}
while \citet{An11} discusses further implications of the separability
assumption.

\subsection{The radius part}

With a separable AD given by equation (\ref{eq:sep}),
equation (\ref{eq:dripn2}) indicates that
(hereafter $x\equiv r^2$),
\begin{equation}
\frd0x\mu\bigl(x^\mu\,\RLI{\mathcal E_0}\Psi{\xi-\frac12}\Nu\bigr)
=\RLI{\mathcal E_0}\Psi{\xi-\frac12}P(\Psi)\cdot
\frd0x\mu[x^\mu R(x)]\ge0
\end{equation}
for $\mu\le\xi$ whereas $\RLI{\mathcal E_0}\Psi{\xi-\frac12}P>0$
for $\xi\ge\frac12$. Therefore, 
\begin{equation}\label{eq:main0}
\frd0x\mu(x^\mu R)\ge0
\qquad(x>0,\,\mu\ge0).
\end{equation}
This is actually equivalent to the condition,
\begin{equation}\label{eq:main1}
R_{(n)}(x)\equiv\frac{\rmd^n[x^nR(x)]}{\rmd x^n}\ge0
\qquad(x>0,\,n=0,1,2,\dotsc),
\end{equation}
which is necessary for the corresponding df to be non-negative
as noted by \citet{An11}.
It is clear that equation (\ref{eq:main0}) implies
equation (\ref{eq:main1}) as the latter is a restriction of
the former for an integer $\mu=n$. The opposite implication follows
Corollary \ref{cor:pdes}):
equation (\ref{eq:main1}) for a positive integer $n$
implies equation (\ref{eq:main0}) for $\mu\in[n-1,n]$ and thus
equation (\ref{eq:main0}) for $\mu\ge0$ follows
equation (\ref{eq:main1}) for all positive integers $n$.

We find more equivalent statements of equation (\ref{eq:main1}).
First equation (\ref{eq:difn}) indicates that
\begin{equation}\label{eq:dxnr}
R_{(n)}(x)
=\frac1{x^{n+1}}\Bigl(x^2\!\frac\rmd{\rmd x}\Bigr)^n\bigl[xR(x)\bigr]
=(-1)^nw^{n+1}\frac{\rmd^n\mathcal R(w)}{\rmd w^n}\biggr\rvert_{w=x^{-1}},
\end{equation}
where
\begin{equation}\label{eq:rlap}
\mathcal R(w)\equiv\frac{R(w^{-1})}w.
\end{equation}
Hence equation (\ref{eq:main1}) is equivalent to
\begin{align}
&\Bigl(x^2\!\frac\rmd{\rmd x}\Bigr)^n\bigl[xR(x)\bigr]\ge0
&&(x>0,\,n=0,1,2,\dotsc),
\\
&(-1)^n\frac{\rmd^n\mathcal R(w)}{\rmd w^n}\ge0
&&(w>0,\,n=0,1,2,\dotsc).
\end{align}
Here the last is also equivalent to saying that the function $\mathcal R(w)$
defined in equation (\ref{eq:rlap}) is a cm function of $w$.
The Bernstein theorem then indicates that
$\mathcal R(w)$ is representable as
the Laplace transformation of a non-negative function.
That is to say, there exists a non-negative function $\phi(t)\ge0$
of $t>0$ such that $\mathcal R(w)=\mathcal L_{t\rightarrow w}[\phi(t)]$.
The inverse Laplace transformation
$\phi(t)=\mathcal L^{-1}_{w\rightarrow t}[\mathcal R(w)]$ may
be found using the Post--Widder formula (\ref{eq:pinv}),
which, thanks to equation (\ref{eq:dxnr}), reduces to
\begin{equation}\label{eq:phit}
\phi(t)=\lim_{n\rightarrow\infty}
\frac1{n!}R_{(n)}\Bigl(\frac tn\Bigr).
\end{equation}
Finally we find another equivalent necessary condition,
\begin{equation}\label{eq:main4}
\lim_{n\rightarrow\infty}
\frac1{n!}\frac{\rmd^n[x^nR(x)]}{\rmd x^n}\biggr\rvert_{x=t/n}\ge0
\qquad(t>0).
\end{equation}
It is obvious that equation (\ref{eq:main1}) implies
equation (\ref{eq:main4}), provided that it converges.
The converse on the other hand
follows the Bernstein theorem and the Post--Widder formula.
However, the conditional
equivalence given the convergence of equation (\ref{eq:phit})
may also be inferred from equation (\ref{eq:dndxnd}). By definition,
equation (\ref{eq:main4}) indicates that there exists
a sufficiently large integer $m$ such that $R_{(n)}(x)\ge0$
for all ${}^\forall n\ge{}^\exists m$ and $x>0$.
Corollary \ref{cor:pdes} then suggests
that $R_{(m-1)}(x)\ge0$ for $x>0$, and equation (\ref{eq:main1})
follows successive arguments on descending subscripts of $R_{(n)}(x)$.

\subsection{The potential part}
\label{sec:potn}

\citet{vHBD11} proved that, given equation (\ref{eq:sep}),
\begin{equation}
P^{(k)}(\Psi)\ge0\qquad(k=0,\dotsc,\flr{\tfrac32-\beta_0})
\end{equation}
where $\beta_0$ is the limit of the anisotropy parameter at the center,
is necessary for the df to be non-negative.
We shall show that this generalizes incorporating fractional
derivatives.

First, we generalize the result of \citet{An11a}
to include arbitrary real order derivatives. This is
trivial since the inverse Abel transform is just
a particular fractional derivative as defined in equation (\ref{eq:frd}).
If the AD is given as equation (\ref{eq:sep}),
equation (\ref{eq:dpirn2}) reduces to
\begin{equation}\label{eq:dpirns}
\frd{\mathcal E_0}\Psi\mu\,
\RLI0x{\xi-\frac12}\Bigl(\frac\Nu{x^{\xi-1/2}}\Bigr)
=\frd{\mathcal E_0}\Psi\mu P\cdot
\RLI0x{\xi-\frac12}\Bigl(\frac R{x^{\xi-1/2}}\Bigr)\ge0,
\end{equation}
for $\mu\le\xi$. Since $R(x)\ge0$ is trivially
necessary, $I^\lambda_x\rvert_0(x^{-\lambda}R)>0$
for $x>0$ and any $\lambda\ge0$ unless $R(x)=0$
\emph{almost everywhere} in $x\equiv r^2\in[0,\infty)$
(Lemma \ref{lem:pos}),
which will not be considered here.
Consequently, equation (\ref{eq:dpirns}) implies that
\begin{equation}\label{eq:pnc}
0<\RLI0x\lambda\Bigl(\frac R{x^\lambda}\Bigr)<\infty
\quad\Longrightarrow\
\frd{\mathcal E_0}\Psi\mu P\ge0
\qquad(\mu\le\lambda+\tfrac12).
\end{equation}
With $\lambda=0$, this
indicates that $\frd{\mathcal E_0}\Psi\mu P\ge0$ for any
$\mu\le\frac12$ -- the condition for $\mu\le0$ is
trivial because $\frd{\mathcal E_0}\Psi{-\lambda}P=
\RLI{\mathcal E_0}\Psi\lambda P$ while $P(\Psi)\ge0$.
For $\lambda>0$ on the other hand, equation (\ref{eq:pnc}) implies that,
if $x^{-\lambda}R(x)\,\rmd x$ is integrable over $x=0$,
then $\frd{\mathcal E_0}\Psi\mu P\ge0$ for any
$\mu\le\lambda+\frac12$ and all accessible $\Psi$ is necessary
for the existence of a non-negative df.
Alternatively, for a fixed $\mu>\frac12$,
equation (\ref{eq:pnc}) suggests
that $\frd{\mathcal E_0}\Psi\mu P\ge0$
is necessary for the df to be non-negative
if there exists ${}^\exists\lambda\ge\mu-\frac12$ such that
$\RLI0x\lambda(x^{-\lambda}R)$ is well-defined.

Equation (\ref{eq:pnc}) however is inconclusive whether
$\frd{\mathcal E_0}\Psi{\frac32-\beta}P\ge0$
is necessary for a non-negative df given
$R(x)\sim x^{-\beta}$ with $\beta<1$ as $x\rightarrow0$ while
this is necessary if we were to extend the result of
\citet{vHBD11}. For this, we first note that
if $f(t)$ is right-continuous at $t=a$,
\begin{equation}\label{eq:dlim}
\lim_{\epsilon\rightarrow0^+}
\epsilon\!\int_a^b\!\frac{h(t)\,\rmd t}{(t-a)^{1-\epsilon}}
=\lim_{t\rightarrow a^+}h(t)=h(a)
\qquad(a<b).
\end{equation}
This applied to the left-hand side of equation (\ref{eq:dpirn2}) reduces to
\begin{subequations}
\begin{gather}
\lim_{\xi\rightarrow(\frac32-\eta)^-}\!
\bigl(\tfrac32-\eta-\xi\bigr)\,
\RLI0x{\xi-\frac12}\Bigl(\frac\Nu{x^{\xi-1/2}}\Bigr)
=\frac{\hat P_\eta(\Psi)}{x^\eta\Gamma(1-\eta)}
\intertext{where $\eta<1$ and}\label{eq:pb}
\hat P_\eta(\Psi)=\lim_{x\rightarrow0^+}x^\eta\Nu(\Psi,x).
\end{gather}
Equation (\ref{eq:dpirn2}) overall then results in the formula,
\begin{equation}\label{eq:nc0}
\frd{\mathcal E_0}\Psi\mu\hat P_\eta(\Psi)
=2^{\frac32-\eta}\pi^\frac32\Gamma(1-\eta)\,
\RLI{\mathcal E_0}\Psi{\frac32-\eta-\mu}
\tilde g_\eta(\Psi)\ge0,
\end{equation}
where
\begin{equation}\label{eq:gb}
\tilde g_\eta(\mathcal E)
=\lim_{L^2\rightarrow0^+}L^{2\eta}\mathcal F(\mathcal E,L^2).
\end{equation}
\end{subequations}
For $\mu<\frac32-\eta$, this is derived with the limit
$\xi\rightarrow(\frac32-\eta)^-$ while maintainting $\mu<\xi<\frac32-\eta$.
For $\mu=\frac32-\eta$ on the other hand, the same limit is taken
with $\mu=\xi$. Therefore, this is valid for
$\mu\le\frac32-\eta$ and $\eta<1$,
provided that $\RLI0x{\xi-\frac12}(x^{\frac12-\xi}\Nu)$
is well-defined for $\xi<\frac32-\eta$ (n.b.,
the integrability of the same for $\xi=\frac32-\eta$ is
actually \emph{not} required for its validity). Here the non-negativity
of equation (\ref{eq:nc0})
follows the non-negativity of $\mathcal F(\mathcal E,L^2)$.
Of particular interests are equation (\ref{eq:nc0}) for $\mu=0$ and
$\frac32-\eta$,
\begin{subequations}
\begin{align}\label{eq:pfromg}
\hat P_\eta(\Psi)
&=2^{\frac32-\eta}\pi^\frac32\Gamma(1-\eta)
\RLI{\mathcal E_0}\Psi{\frac32-\eta}\tilde g_{\eta}(\Psi);
\\\label{eq:cuinv}
\tilde g_\eta(\Psi)
&=\frac{\frd{\mathcal E_0}\Psi{\frac32-\eta}\hat P_\eta(\Psi)}
{2^{\frac32-\eta}\pi^\frac32\Gamma(1-\eta)},
\end{align}
\end{subequations}
that is, explicit formulae for $\hat P_\eta(\Psi)$
and $\tilde g_\eta(\Psi)$ from each other.

For a separable AD given as in equation (\ref{eq:sep}), we have
\begin{equation}
\hat P_\eta(\Psi)=\bar R_\eta P(\Psi)\,;\qquad
\bar R_\eta=\lim_{x\rightarrow0^+}x^\eta R(x),
\end{equation}
Therefore, equation (\ref{eq:nc0}) indicates that
\begin{equation}
0<\bar R<\infty
\quad\Longrightarrow\
\frd{\mathcal E_0}\Psi\mu P\ge0
\qquad(\mu\le\tfrac32-\eta).
\end{equation}
That is to say, if there exists ${}^\exists\eta<1$ such that
$\bar R_\eta$ is a (non-zero) positive finite constant,
then $\frd{\mathcal E_0}\Psi\mu P\ge0$
for any ${}^\forall\mu\le\frac32-{}^\exists\eta$.
This actually encompasses equation (\ref{eq:pnc}), which is seen 
as follows: If $\bar R_\eta$ is non-zero finite for $\eta<1$,
then we basically find that $R\sim x^{-\eta}$
as $x\rightarrow0$. Hence $\RLI0x\lambda(x^{-\lambda}R)$
converges for $\lambda<1-\eta$, and so
if $\mu\le\lambda+\frac12$, then $\mu<\frac32-\eta$.

For example, with a constant anisotropy system given by 
\begin{equation}\label{eq:cbeta}
R(x)=x^{-\beta};
\qquad
\bar R_\beta=1
\qquad(\beta\le1)
\end{equation}
the convergence condition reduces to
\[
\RLI0x\lambda\Bigl(\frac R{x^\lambda}\Bigr)
=\frac1{\Gamma(\lambda)}\!\int_0^x\!
\frac{(x-s)^{\lambda-1}\,\rmd s}{s^{\lambda+\beta}}
=\frac{\Gamma(1-\beta-\lambda)}{x^\beta\Gamma(1-\beta)}<\infty,
\]
which converges if $0\le\lambda<1-\beta$.
It follows that equation (\ref{eq:pnc}) indicates that 
$\frd{\mathcal E_0}\Psi\mu P(\Psi)\ge0$
for $\mu\le\lambda+\frac12<\frac32-\beta$ is necessary
for the df to be non-negative
whereas equation (\ref{eq:nc0}) suggests the same
for $\mu\le\frac32-\beta$.

\section{Sufficient conditions for
phase-space consistency
in terms of separable augmented densities}
\label{sec:suf}

Recently, \citet{vH11} derived the necessary and sufficient
condition for the df with $\mathcal E_0=0$
to be non-negative, expressed in terms of
the integro-differential constraints on the AD. They achieved this
by reducing the problem to the Hausdorff moment problem,
according to which the df is non-negative if and only if
the moment sequence in equation (\ref{eq:moms}) is
a completely monotone sequence.\footnote{A sequence
$(a_0,a_1,a_2,\dotsc)$ is completely monotone if and only if
$(-1)^k\Delta^ka_j\ge0$ for any non-negative integer pairs $k$ and $j$.
Here $\Delta$ is the forward finite difference operator defined such that
$\Delta^{k+1}a_j=\Delta^ka_{j+1}-\Delta^ka_j$ and $\Delta^0a_j=a_j$.}
Since the moment sequence are
generated by the AD using equation (\ref{eq:msad}),
the monotone sequence condition is expressible in terms of
finite differences of integro-differential operations on the AD.

With a separable AD, they have derived a simpler
sufficient (but not necessary) condition given as a union of
conditions, each of which only involves the potential
or the radius part separately but not together.
Here we derive an alternative sufficient condition for a separable AD
to be resulted from a non-negative df, which turns out to be equivalent
to that of \citet{vH11}.
The derivation here is based on the properties of cm functions
and also uses the Laplace transform extensively. In this
section, we only consider the case that $\mathcal E_0=0$
and $\bar L^2=2r^2\Psi$, that is, the df has a compact support
and $\mathcal F(\mathcal E<0,L^2)=0$.

\subsection{Inversion of a separable augmented density
for the distribution function}

As it has been shown by \cite[see also \citealp{An11a}]{HQ93},
inverting equation (\ref{eq:dist}) for $\mathcal F(\mathcal E,L^2)$
is formally equivalent to recovering the two-integral even df,
$\mathcal F^+(\mathcal E,J_z^2)$ from the axisymmetric density
$\nu[\Psi(R^2,z^2),R^2]$. The findings of the preceding section
together with the inversion of \citet{Ly62} who utilized the Laplace
transform for the latter problem suggest that
the function $\phi(t)$ defined by equation (\ref{eq:phit}) must
be directly related to the underlying df, $\mathcal F(\mathcal E,L^2)$.
We investigate this connection in the following.

Following \citet{Ly62}, we apply the Laplace transform on $\Psi$ to
equation (\ref{eq:dist}),
\begin{subequations}
\begin{multline}
\laplace\Psi s\bigl[\Nu(\Psi,r^2)\bigr]
=\int_0^\infty\!\rmd\Psi\,\rme^{-s\Psi}\Nu(\Psi,r^2)
\\=\frac{2\pi}{r^2}\!
\iint\limits_{\mathcal E\ge0,L^2\ge0}\!\rmd\mathcal E\,\rmd L^2
\mathcal F(\mathcal E,L^2)\!\int_0^\infty\!\rmd\Psi\,\rme^{-s\Psi}
\frac{\Theta(\mathcal K)}{\sqrt{\abs{\mathcal K}}}.
\end{multline}
The inner integral in the right-hand side reduces to
\begin{multline}
\int_0^\infty\!\rmd\Psi\,\rme^{-s\Psi}
\frac{\Theta(\mathcal K)}{\sqrt{\abs{\mathcal K}}}
=\exp\biggl\lgroup-s\mathcal E-\frac{sL^2}{2r^2}\biggr\rgroup
\int_0^\infty\!\frac{\rmd\mathcal K}2\,
\rme^{-\frac{s\mathcal K}2}\mathcal K^{-\frac12}
\\
=\sqrt{\frac\pi{2s}}\,\rme^{-s\mathcal E}\,\rme^{-\frac{sL^2}{2r^2}},
\end{multline}
and consequently we find that
%
\begin{equation}
\laplace\Psi s[\Nu]
=\frac{\sqrt2\pi^\frac32}{s^\frac12r^2}\!
\int_0^\infty\!\rmd L^2\rme^{-\frac{sL^2}{2r^2}}\!
\int_0^\infty\!\rmd\mathcal E\,\rme^{-s\mathcal E}
\mathcal F(\mathcal E,L^2).
\end{equation}
\end{subequations}
Substituting variables,
$t=\frac12sL^2$ and $w=r^{-2}$, this reduces to
%
\begin{equation}\label{eq:laps}
\Bigl(\frac s{2\pi}\Bigr)^\frac32
\laplace\Psi s\biggl[\frac{\Nu(\Psi,w^{-1})}w\biggr]
=\laplace tw\biggl[\int_0^\infty\!\rmd\mathcal E\,\rme^{-s\mathcal E}
\mathcal F\Bigl(\mathcal E,\frac{2t}s\Bigr)\biggr].
\end{equation}

If the AD is separable (eq.~\ref{eq:sep}),
then $w^{-1}\Nu(\Psi,w^{-1})=P(\Psi)\mathcal R(w)$ where
$\mathcal R(w)$ is as defined in equation (\ref{eq:rlap})
and so the left-hand side becomes
\begin{equation}\label{eq:lapsep}
\frac{s^\frac32\mathcal P(s)}{(2\pi)^{\frac32}}\mathcal R(w)
=\laplace tw\biggl[
\frac{s^\frac32\mathcal P(s)}{(2\pi)^{\frac32}}\phi(t)\biggr].
\end{equation}
Here $\mathcal P(s)$ is the Laplace transformation of $P(\Psi$),
\begin{equation}\label{eq:pdef}
\mathcal P(s)\equiv\laplace\Psi s[P(\Psi)]
=\int_0^\infty\!\rmd\Psi\,\rme^{-s\Psi}P(\Psi).
\end{equation}
We have also used $\mathcal R(w)=\mathcal L_{t\rightarrow w}[\phi(t)]$.
Given that the inverse Laplace transformation is unique,
equating the right-hand sides of equations (\ref{eq:laps})
and (\ref{eq:lapsep}) results in
\begin{subequations}
\begin{equation}
\frac{s^\frac32\mathcal P(s)\phi(t)}{(2\pi)^\frac32}
=\int_0^\infty\!\rmd\mathcal E\,\rme^{-s\mathcal E}
\mathcal F\Bigl(\mathcal E,\frac{2t}s\Bigr).
\end{equation}
Finally reinstating $t=\frac12sL^2$ leads to
\begin{equation}\label{eq:elap}
\frac{s^\frac32\mathcal P(s)}{(2\pi)^\frac32}
\phi\Bigl(\frac{sL^2}2\Bigr)
=\laplace{\mathcal E}s\bigl[\mathcal F(\mathcal E,L^2)\bigr].
\end{equation}
\end{subequations}
The df is then recovered via the inverse Laplace transform,
\begin{equation}\label{eq:lapdf}
\mathcal F(\mathcal E,L^2)=
\ilaplace s{\mathcal E}
\biggl[\frac{s^\frac32\mathcal P(s)}{(2\pi)^\frac32}
\phi\Bigl(\frac{sL^2}2\Bigr)\biggr].
\end{equation}

\subsection{Sufficient condition on a separable
augmented density}

According to the Bernstein theorem, the df in
equation (\ref{eq:lapdf}) is non-negative if and only if
the left-hand side of equation (\ref{eq:elap}) is a cm function of $s>0$
for all accessible values of $L^2$. However $\mathcal P(s)$ defined
in equation (\ref{eq:pdef}) is already cm since $P(\Psi)\ge0$.
Hence Lemma \ref{lem:cm} suggests that $s^\frac32\phi(sL^2/2)$ is
a cm function of $s>0$ for any $L^2\ge0$ is in fact a sufficient
condition for the non-negativity of the df. Equivalently, since
\begin{equation}
\frac{\rmd^n[t^{\frac32}\phi(t)]}{\rmd t^n}\biggr\rvert_{t=sL^2/2}
=\Bigl(\frac{L^2}2\Bigr)^{\frac32-n}\!
\frac{\rmd^n}{\rmd s^n}
\biggl[s^{\frac32}\phi\Bigl(\frac{sL^2}2\Bigr)\biggr],
\end{equation}
the condition is also equivalent to that $t^\frac32\phi(t)$ is cm.
Unfortunately, this is too severe to be physically relevant\footnote{If
the Laplace transform of $\phi(t)$ exists, then $\phi(t)$ cannot
diverges faster than $t^{-1}$ as $t\rightarrow0$. Consequently,
$\lim_{t\rightarrow0}t^{3/2}\phi(t)\rightarrow0$ and thus
$t^{3/2}\phi(t)$ cannot be cm because the limit suggests that
$t^{3/2}\phi(t)$ should be negative or increasing in some interval
$t\in(0,t_0)$ where ${}^\exists t_0>0$.},
which may be inferred from the constant anisotropy model 
given by equation (\ref{eq:cbeta}). With this model, we find
for $\beta<1$
\begin{equation}\label{eq:cbetap}
R_{(n)}(x)=\frac{(1-\beta)_n^+}{x^\beta}\,;\qquad
\phi(t)=\frac1{t^\beta\Gamma(1-\beta)}
\end{equation}
where we have used
\begin{equation}\label{c5}
\lim_{n\rightarrow\infty}\frac{(1+z)_n^+}{n!n^z}=\frac1{\Gamma(1+z)}
\end{equation}
to find $\phi(t)$ using equation (\ref{eq:phit}).
The condition thus reduces to
\begin{equation}
\frac{(\beta-\frac32)_n^+}
{\Gamma(1-\beta)}\frac1{t^{\beta+n-\frac32}}\ge0
\qquad(t>0,\,n=0,1,2,\dotsc),
\end{equation}
for $t>0$ and all non-negative integers $n$,
which cannot be satisfied for any constant $\beta<1$.

Nevertheless, the preceding discussion extends to yield
useful sufficient conditions. That is,
for any fixed $\lambda$, the conditions that
\begin{align}
&(-1)^n\frac{\rmd^n[s^\lambda\mathcal P(s)]}{\rmd s^n}\ge0
&&(s>0,\,n=0,1,2,\dotsc),
\label{eq:pcon0}\\\label{eq:rcon}
&(-1)^n\frac{\rmd^n[t^{\frac32-\lambda}\phi(t)]}{\rmd t^n}\ge0
&&(t>0,\,n=0,1,2,\dotsc)
\end{align}
are jointly sufficient to imply equation (\ref{eq:elap})
being cm and consequently the non-negativity
of the df. With increasing $\lambda$,
the constraint in equation (\ref{eq:pcon0}) tightens whereas
the condition in equation (\ref{eq:rcon}) becomes strictly weaker.
In other words, with a larger $\lambda$,
the smaller subset of functions $P(\Psi)$ will lead to
$s^\lambda\mathcal P(s)$ being cm. At the same time
if $\phi(t)$ satisfies
equation (\ref{eq:rcon}) for a fixed $\lambda=\lambda_0$,
the same condition for any larger $\lambda\ge\lambda_0$
automatically holds. Both of these are easily inferred from
Corollary \ref{cor:cmp}.

\subsubsection{the condition on $R(x)$ equivalent to eq.~(\ref{eq:rcon})}

Both conditions can also be translated into the direct constraints on
the behaviors of $P(\Psi)$ and $R(r^2)$. For the radius part,
we use that $\phi(t)$ may be given by equation (\ref{eq:phit}).
Note that the existence of $\phi(t)$ and
the validity of equation (\ref{eq:phit}) as well as
its non-negativity, that is, $\phi(t)\ge0$ for ${}^\forall t>0$
are all necessary. Substituting equation (\ref{eq:phit}) into
the left-hand side of equation (\ref{eq:rcon}) results in
\begin{multline}
(-1)^n\frac{\rmd^n[t^{\frac32-\lambda}\phi(t)]}{\rmd t^n}
=\lim_{k\rightarrow\infty}\frac{(-1)^n}{k!}\frac{\rmd^n}{\rmd t^n}
\biggl[t^{\frac32-\lambda}R_{(k)}\Bigl(\frac tk\Bigr)\biggr]
\\=\lim_{k\rightarrow\infty}\frac{(-1)^n}{k!k^{n+\lambda-\frac32}}
\frac{\rmd^n[x^{\frac32-\lambda}R_{(k)}(x)]}{\rmd x^n}
\biggr\rvert_{x=t/k}.
\end{multline}
Consequently, provided that this limit converges,
equation (\ref{eq:rcon}) is equivalent to
insisting that there exists a sufficiently large integer $m$
such that, for all integers ${}^\forall k\ge{}^\exists m$,
\begin{equation}\label{eq:rcon0}
(-1)^n\frac{\rmd^n}{\rmd x^n}
\biggl\lbrace x^{\frac32-\lambda}\frac{\rmd^k[x^kR(x)]}{\rmd x^k}
\biggr\rbrace\ge0
\qquad(x>0,\,n=0,1,2,\dotsc).
\end{equation}
That is to say, $x^{\frac32-\lambda}R_{(k)}(x)$ being cm
for all sufficiently large integers $k$ is necessary and sufficient
for $\phi(t)$ derived from the same $R(x)$
to satisfy equation (\ref{eq:rcon}), provided that the limit
converges.
In fact, equation (\ref{eq:rcon}) is equivalent to
equation (\ref{eq:rcon0}) for not only all sufficiently
large integers but also all non-negative integers $k$,
thanks to Theorem \ref{th:cmdes}, which indicates that
$x^{\frac32-\lambda}R_{(m+1)}(x)$ being cm implies
$x^{\frac32-\lambda}R_{(m)}(x)$ being also cm.
Succesive arguments with descending $k$ then
establish that $x^{\frac32-\lambda}R_{(k)}(x)$ being cm for
all sufficiently large integers $k$ implies that
$x^{\frac32-\lambda}R_{(k)}(x)$ is a cm function for
all non-negative integers $k$ (the opposite implication is trivial).
Note that the condition as stated in equation (\ref{eq:rcon0})
for all non-negative integers $k$ has already been noted by \citet{vH11}.

\subsubsection{the condition on $P(\Psi)$ equivalent to
eq.~(\ref{eq:pcon0})}

The explicit constraints on $P(\Psi)$ resulted from
equation (\ref{eq:pcon0})
are derived by means of fractional calculus.
We first find,
from equations (\ref{eq:lapderiv}) and (\ref{eq:lapab}), that
(n.b., $\RLI0\Psi\xi P(0)=0$ for $\xi>0$ from Corollary \ref{eq:if0})
\begin{align}\label{eq:slp}
&s^\lambda\mathcal P(s)
=\frac{s^{\mu+1}}{s^{1-\delta}}\laplace\Psi s[P(\Psi)]
=s^{\mu+1}\!\laplace\Psi s\bigl[\RLI0\Psi{\!1-\delta}P(\Psi)\bigr]
\\\nonumber&\quad
=\laplace\Psi s\bigl[\frd0\Psi\lambda P(\Psi)\bigr]
+{\textstyle\sum_{j=1}^\mu\!s^{j-1}\,\frd0\Psi{\lambda-j}P(0)}
+s^\mu\,\cancel{\RLI0\Psi{\!1-\delta}P(0)}
\end{align}
where $\mu=\flr\lambda$ and $0\le\delta=\lambda-\mu<1$.
This indicates that
\begin{gather}
\frd0\Psi\lambda P(\Psi)\ge0
\qquad(\Psi>0)
\label{eq:cons2}\\\label{eq:cons1}
\cancelto0{\RLI0\Psi{\!1-\delta}P(0)}=
\frd0\Psi\delta P(0)=\dotsb=
\frd0\Psi{\lambda-1}P(0)=0,
\end{gather}
for $\lambda\ge0$
is a sufficient condition for $s^\lambda\mathcal P(s)$ to be cm.
Note, provided that $P(\Psi)$ is right-continuous at $\Psi=0$,
that $\RLI0\Psi{\!1-\delta}P(0)=0$ (cf., eq.~\ref{eq:if0}),
which will thus be taken as granted. Consequently,
equation (\ref{eq:cons1}) for $0\le\lambda<1$ is essentially
an empty condition.
For $\lambda=0$, equation (\ref{eq:cons2}) reduces to $P(\Psi)\ge0$.
For a positive integer $\lambda=m+1$
on the other hand, the condition is equivalent to
\begin{equation}\label{eq:pcon}
P^{(m+1)}(\Psi)\ge0\qquad
\&\qquad
P(0)=\dotsb=P^{(m)}(0)=0.
\end{equation}
For $0<\delta<1$, the condition in
equation (\ref{eq:cons1}) may also be replaced with the same
boundary condition as equation (\ref{eq:pcon}). In particular,
thanks to Lemma \ref{th:p0},
$P^{(0)}(0)=\dotsb=P^{(n)}(0)=0$
implies $\frd0\Psi{n+\delta}P(0)=0$ for $0<\delta<1$.
Consequently, it follows that for $\lambda\ge1$,
\addtocounter{equation}{-2}\begin{subequations}
\begin{equation}\label{eq:cons1a}
P^{(0)}(0)=\dotsb=P^{(\flr\lambda-1)}(0)
\end{equation}\end{subequations}
actually implies equation (\ref{eq:cons1}) -- if $\delta=0$, they
are identical. Therefore, equations (\ref{eq:cons2}) and
(\ref{eq:cons1a}) together also consist in a sufficient condition
for $s^\lambda\mathcal P(s)$ to be cm at a fixed $\lambda$.
The condition expressed with equation (\ref{eq:cons1a}) is useful
because equation (\ref{eq:frda}) then indicates that
equation (\ref{eq:cons2}) is equivalent to
\addtocounter{equation}1
\begin{equation}
\frd0\Psi\lambda P
=\frac1{\Gamma(1-\delta)}\frac{\rmd^{1+\mu-n}}{\rmd\Psi^{1+\mu-n}}\!
\int_0^\Psi\!\frac{P^{(n)}(Q)\,\rmd Q}{(\Psi-Q)^\delta}\ge0
\end{equation}
%
where $n$ is any non-negative integer not greater than $\lambda$.

Again, the joint condition of
equations (\ref{eq:cons2}) and (\ref{eq:cons1a})
becomes strictly stronger as $\lambda$
increases in accordance with the restriction
on the complete monotonicity of $s^\lambda\mathcal P(s)$.
This is seen using equation (\ref{eq:difcomb})
for $0\le\epsilon\le\lambda$ under the condition
of equation (\ref{eq:cons1}) or (\ref{eq:cons1a}),
\[
\RLI0\Psi\epsilon\frd0\Psi\lambda P
=\frd0\Psi{\lambda-\epsilon}P
-\cancelto0{\sum_{k=1}^\mu
\frac{(\epsilon)_k^-\,\frd0\Psi{\lambda-k}P(0)}
{\Psi^{k-\epsilon}\Gamma(1+\epsilon)}}.
\]
That is, $\frd0\Psi\lambda P(\Psi)\ge0$
implies $\frd0\Psi\xi P(\Psi)\ge0$
for $0\le\xi\le\lambda$. The similar
implications of equation (\ref{eq:cons1a}) with descending
$\lambda$ are trivial.

\subsection{the constant anisotropy model}
\label{sec:cona}

As an illustrative example, let us
consider the constant anisotropy model with $\beta<1$ (see
Appendix \ref{app:b1} for the $\beta=1$ case) given by
in equation (\ref{eq:cbeta}). Given equation (\ref{eq:cbetap}),
equations (\ref{eq:rcon}) and (\ref{eq:rcon0}) now reduce to
\begin{subequations}
\begin{gather}
(-1)^n\frac{\rmd^n[t^{\frac32-\lambda}\phi(t)]}{\rmd t^n}=
\frac{(\beta+\lambda-\frac32)_n^+}{\Gamma(1-\beta)}
\frac1{t^{\beta+n+\lambda-3/2}}\ge0\,;
\\
(-1)^n\frac{\rmd^n[x^{\frac32-\lambda}R_{(k)}(x)]}{\rmd x^n}=
(1-\beta)_k^+
\frac{(\beta+\lambda-\frac32)_n^+}{x^{\beta+n+\lambda-3/2}}
\ge0.
\end{gather}
\end{subequations}
Thus $\beta+\lambda\ge\frac32$ and $\beta<1$ is
sufficient for these to be satisfied. If $\lambda=m+1$ is
a positive integer and $\beta\ge\frac12-m$,
then equation (\ref{eq:pcon}) is sufficient for the existence
of a non-negative df \cite[cf.,][]{CM10}.
Our result furthermore implies for any 
\emph{real} $\lambda>\frac12$ that if $\frac32-\lambda\le\beta<1$,
then equations (\ref{eq:cons2}) and (\ref{eq:cons1a})
constitutes a sufficient condition.

With a fixed $\beta<1$, this indicates that, if there exists
${}^\exists\lambda\ge\frac32-\beta$
such that equations (\ref{eq:cons2}) and (\ref{eq:cons1a})
hold, then the corresponding $P(\Psi)$ guarantees the phase-space
consistency. This also implies
$\frd0\Psi\xi_\Psi P\ge0$ for $0\le{}^\forall\xi\le{}^\exists\lambda$.
while Sect.~\ref{sec:potn} indicates that,
for the same system, $\frd0\Psi\xi_\Psi P\ge0$
for ${}^\forall\mu\le\frac32-\beta$ 
is necessary for the df to be non-negative.
It follows that, if $\Nu(\Psi,r^2)=r^{-2\beta}P(\Psi)$, then
$\frd0\Psi{\frac32-\beta}P\ge0$ 
is the necessary \emph{and} sufficient condition
for the non-negative df. In fact, here 
$P(\Psi)=\hat P_\beta(\Psi)$ and
$\mathcal F(\mathcal E,L^2)=\tilde g_\beta(\mathcal E)L^{-2\beta}$
where $\hat P_\beta(\Psi)$ and $\tilde g_\beta(\mathcal E)$ are as
defined in equations (\ref{eq:pb}) and (\ref{eq:gb}) with $\eta=\beta$,
and so equation (\ref{eq:cuinv}) results in the inversion formula,
\begin{equation}
\mathcal F(\mathcal E,L^2)
=\frac{\frd0{\mathcal E}{\frac32-\beta}P(\mathcal E)}
{2^{\frac32-\beta}\pi^\frac32\Gamma(1-\beta)L^{2\beta}}
\quad\Longleftarrow\
\Nu(\Psi,r^2)=\frac{P(\Psi)}{r^{2\beta}}.
\end{equation}
This is simply the generalized \citeauthor{Ed16}\footnote{Sir
Arthur Stanley Eddington (1882-1944)} inversion formula
\cite[see e.g.,][]{EA06} for constant anisotropy systems.
That $\frd0\Psi{\frac32-\beta}P(\Psi)\ge0$ is
necessary and sufficient for the existence of a non-negative df
is a trivial consequence of the inversion formula.

\section{Family of monotonic anisotropy parameters}

Consider the anisotropy parameterized to be \citep[c.f.,][]{BvH07,An11}
\begin{subequations}
\begin{equation}
\beta(r)
=\frac{\beta_1r_\mathrm a^{2s}+\beta_2r^{2s}}
{r_\mathrm a^{2s}+r^{2s}}
\qquad(s>0,\,r_\mathrm a>0).
\end{equation}
If the spherical system is characterized by a separable AD
as in equation (\ref{eq:sep}), this follows the radial function
(see eq.~\ref{eq:beta})
\begin{equation}\label{eq:rfun}\begin{split}
R(x)&=x^{-\beta_1}(1+x^s)^{-\zeta}
\qquad\text{where }\ s\zeta=\beta_2-\beta_1;
\\\mathcal R(w)&=w^{-1}R(w^{-1})=w^{\beta_2-1}(1+w^s)^{-\zeta}
\end{split}\end{equation}
\end{subequations}
where $x=r^2/r_\mathrm a^2$ (i.e., $r_\mathrm a=1$),
which does not affect the following discussion.
Note $R_{(1)}(x)\ge0$ for $x>0$
restricts $\beta_1,\beta_2\le1$.
\citet{An11} also provides an elementary proof that if $0<s\le1$
and $\beta_1,\beta_2\le1$, equation (\ref{eq:rfun})
satisfies equation (\ref{eq:main1}). The same is deduced from
the complete monotonicity of $\mathcal R(w)$
for $0<s\le 1$ and $\beta_1,\beta_2\le1$ (Corollary \ref{cor:qcm}), too.

The situation for $s>1$
however is inconclusive: on one hand, if $\beta_2=1>\beta_1$, it
is easy to show that $\mathcal R''(w)<0$ for $0<w^s<(s-1)/(2-\beta_1)$
and so the condition fails for $s>1$ whereas \citet{An11}
on the other hand has found that the condition is met for all $s>0$ if
$-\zeta=(\beta_1-\beta_2)/s$ is non-negative integer. It appears
that for a fixed $s>1$, there exists a \emph{proper} subset of parameter
combinations $\beta_1,\beta_2\le1$ that satisfies the necessary condition
of equation (\ref{eq:main1}), but we have not been able to establish
the concrete criteria. The necessary
condition on the potential part discussed in Sect.~\ref{sec:potn}
on the other hand is straightforward. That is,
given $R(x)$ of equation (\ref{eq:rfun}), the potential part
$P(\Psi)$ must satisfy
$\frd{\mathcal E_0}\Psi\lambda P\ge0$
for any ${}^\forall\lambda\le\frac32-\beta_1$ in order for the df to be
non-negative. Here also note $\beta_1\le1$ and thus 
$\frd{\mathcal E_0}\Psi\lambda P\ge0$ for any $\lambda\le\frac12$.

For $0\le x<1$, the binomial expansion of equation (\ref{eq:rfun})
and the subsequent term-by-term differentiation indicate that
\begin{equation}\begin{split}
R&(x)=\sum\nolimits_{k=0}^\infty\frac{(-1)^k(\zeta)_k^+}{k!}x^{sk-\beta_1};
\\R&_{(n)}(x)
=\sum\nolimits_{k=0}^\infty\frac{(-1)^k(\zeta)_k^+}{k!}\,
(1-\beta_1+sk)_n^+x^{sk-\beta_1}.
\end{split}\label{c3}\end{equation}
It follows equations (\ref{eq:phit}) and (\ref{c5}) that
\begin{equation}\label{infs}\begin{split}
\phi(t)&=\sum\nolimits_{k=0}^\infty\frac{(-1)^k(\zeta)_k^+}{k!}\,
t^{sk-\beta_1}
\lim_{n\rightarrow\infty}\frac{(1-\beta_1+sk)_n^+}{n!n^{sk-\beta_1}}
\\&=\sum\nolimits_{k=0}^\infty
\frac{(-1)^k(\zeta)_k^+t^{sk-\beta_1}}{k!\Gamma(1-\beta_1+sk)}
=t^{-\beta_1}E_{s,1-\beta_1}^\zeta({-t^s}),
\end{split}\end{equation}
where $E_{p,b}^\lambda(z)$ is the extended
generalization of the Mittag-Leffler\footnote{Magnus Gustaf (G\"osta)
Mittag-Leffler (1846-1927)} function $E_{p,b}^\lambda$
introduced by \cite[see also \citealt{HMS}]{Pr71}.
Although the derivation here is essentially formal (see
Appendix \ref{sec:app} for proper treatments) as in that
we have not properly considered the issue of the convergence,
the result is in fact valid given that $\beta_1<1$ (for $\beta_1=1$,
see Appendix~\ref{app:b1}) as is found in equation (\ref{eq:gmllt}).
Next we briefly detour to examine properties of generalized
Mittag-Leffler functions necessary to derive sufficient
conditions in Sect.~\ref{sec:suf} for the phase-space consistency
given that the radial part is given by equation (\ref{eq:rfun}).

\subsection{Generalized Mittag-Leffler function}

\begin{definition}
Let us consider a particular generalized hypergeometric function
defined to be
\begin{equation}\label{eq:gml}
E^\lambda_{p,b}(z)\equiv\sum_{k=0}^\infty
\frac{(\lambda)_k^+}{\Gamma(pk+b)}\frac{z^k}{k!}
\qquad(p>0).
\end{equation}
\end{definition}
Note that the Stirling\footnote{James Stirling (1692-1770)}
approximation suggests
\[\lim_{n\rightarrow\infty}\frac{\Gamma(n)}{\Gamma(n+x)}
=\lim_{n\rightarrow\infty}\frac1{(n+x)^x}
=\begin{cases}
0&(x>0)\\1&(x=0)\\\infty&(x<0),
\end{cases}\]
and so the ratio test for equation (\ref{eq:gml}) with $p>0$
\[\lim_{k\rightarrow\infty}\left\lvert\frac{\lambda+k}{k+1}
\frac{z\Gamma(pk+b)}{\Gamma(pk+b+p)}\right\rvert=0\]
indicates that the infinite series for $p>0$ absolutely converges
for all (finite) $z$. It follows that $E_{p,b}^\lambda(z)$ with
$p>0$ is an entire function of $z$. This is indeed a generalization
of the Mittag-Leffler function since
\[E^1_{p,b}(z)=E_{p,b}(z)\,;\qquad
E^1_{p,1}(z)=E_{p,1}(z)=E_p(z)\]
where $E_p(z)$ and $E_{p,b}(z)$ are the classical Mittag-Leffler
function and its generalization by \citet{Wi05}. If $p=1$ on the other
hand, this reduces to the Kummer\footnote{Ernst Eduard Kummer (1810-1893)}
confluent huypergeometric function of the first kind, that is,
\[E^\lambda_{1,b}(z)={}_1\tilde F_1(\lambda;b;z)
=\frac{{}_1F_1(\lambda;b;z)}{\Gamma(b)}.\]
Finally $E_{p,b}^\lambda(z)$ with $\lambda\ne0$ is also a particular
case of the \citeauthor{Wr35}\footnote{Sir Edward Maitland Wright
(1906-2005)} generalized hypergeometric function ${}_1\varPsi_1$,
\begin{subequations}
\begin{equation}
E_{p,b}^\lambda(z)
=\frac1{\Gamma(\lambda)}\,{}_1\varPsi_1\!\left[\begin{array}c
(\lambda,1);\\(b,p);\end{array}z\right],
\end{equation}
and also the \citeauthor{Fx61}\footnote{Charles Fox (1897-1977)} H-function,
\begin{equation}\label{eq:fox}
E_{p,b}^\lambda(z)
=\frac1{\Gamma(\lambda)}\,
H^{1,1}_{1,2}\!\left\lgroup{-z}\,\vrule\begin{array}c
\lbrace1-\lambda,1\rbrace\\\lbrace0,1\rbrace,\lbrace1-b,p\rbrace
\end{array}\right\rgroup.
\end{equation}
\end{subequations}

Next, the term-by-term integration indicates that for $\lambda>0$
\begin{equation}\label{eq:gmlfrd}
E^\lambda_{p,b}
(\pm z)=\frac1{\Gamma(\lambda)}
\frd0z{\lambda-1}\bigl[z^{\lambda-1}E_{p,b}(\pm z)\bigr]
\end{equation}
where we have used Lemma \ref{lem:frd}. Together with 
the integral representation of the Mittag-Leffler function,
\[E_{p,b}(\pm z)=\frac1{2\pi\mathrm i}
\int_\Omega\frac{t^{p-b}\rme^t\rmd t}{t^p\mp z},\]
and using (c.f., Lemma \ref{lem:frd} and the binomial expansion)
\[\frd0z{\lambda-1}\Bigl(\frac{z^{\lambda-1}}{t^p\mp z}\Bigr)
=\sum_{k=0}^\infty\frac{\Gamma(k+\lambda)(\pm z)^k}{k!t^{pk+p}}
=\frac{t^{p\lambda-p}\Gamma(\lambda)}{(t^p\mp z)^\lambda},\]
this then leads to the integral representation
\begin{equation}\label{eq:gmlint}
E^\lambda_{p,b}(\pm z)=\frac1{2\pi\mathrm i}
\int_\Omega\frac{t^{p\lambda-b}\rme^t\rmd t}{(t^p\mp z)^\lambda}.
\end{equation}
Here the integral loop $\Omega$ is the same as usual for the
Mittag-Leffler function, that is, it starts and ends at `$-\infty$',
and loops around the circle $\abs t=\abs z^{1/p}$ in positive sense.
This may also be independently proven using the Hankel\footnote{Hermann
Hankel (1839-1873)}-loop integral
\[\frac1{\Gamma(z)}=\frac1{2\pi\mathrm i}
\int_{-\infty}^{(0+)}\frac{\rme^t\rmd t}{t^z}\]
for the reciprocal gamma function
(and also using the binomial expansion),
similarly to the classical case.
Equation (\ref{eq:gmlint}) implies the
\emph{asymptotic expansion} as $z\rightarrow+\infty$,
\begin{equation}\label{eq:gmlasym}\begin{split}
E^\lambda_{p,b}(z^p)&\sim
\frac{z^{\lambda-b}\rme^z}{p^\lambda\Gamma(\lambda)},
\\E^\lambda_{p,b}(-z)&\simeq
\sum_{k=0}^\infty\frac{(-1)^k(\lambda)_k^+}
{k!\Gamma(b-p\lambda-pk)z^{\lambda+k}}
\sim\frac1{\Gamma(b-p\lambda)}\frac1{z^\lambda}.
\end{split}\end{equation}

If $\xi=-\lambda$ is a non-negative integer,
the series in equation (\ref{eq:gml})
terminates after the finite number of terms
and thus reduces to a polynomial on $z$
-- in particular, $E_{p,b}^0=1/\Gamma(b)$ is constant.
In general, if $\xi=-\lambda\ge0$, an alternative expression with
the Fox H-function is also derived by separating
the sum up to $k=\flr\xi$. That is, 
equation (\ref{eq:gml}) is alternatively given by
\begin{multline*}
E^{-\xi}_{p,b}(z)
=\sum_{k=0}^{\flr\xi}\binom\xi k\,\frac{(-z)^k}{\Gamma(pk+b)}
\\+(\delta)_{\flr\xi+1}^+(-z)^{\flr\xi+1}
\sum_{k=0}^\infty
\frac{(1-\delta)_k^+}{\Gamma(pk+\hat b_{\flr\xi,p})}
\frac{z^k}{(k+\flr\xi+1)!}
\end{multline*}
where $\hat b_{\mu,p}=p(\mu+1)+b$ and $\delta=\xi-\flr\xi$.
For $0<\delta<1$, the last infinite sum here results in
\[\begin{split}
\Gamma(1-\delta)&\sum_{k=\mu+1}^\infty
\frac{(1-\delta)_k^+}{\Gamma\bigl(pk+\hat b_{\mu,p}\bigr)}
\frac{z^k}{(k+\mu+1)!}
\\&={}_2\varPsi_2\!\left[\begin{array}c
(1-\delta,1),(1,1);\\\bigl(\hat b_{\mu,p},p\bigr),(\mu+2,1);
\end{array}z\right]
\\&=H^{1,2}_{2,3}\!\left\lgroup{-z}\,\vrule\begin{array}c
\lbrace\delta,1\rbrace,\lbrace0,1\rbrace
\\\lbrace0,1\rbrace,\lbrace{-\mu-1},1\rbrace,
\bigl\lbrace1-\hat b_{\mu,p},p\bigr\rbrace
\end{array}\right\rgroup.
\end{split}\]
The convergent integration path for the last Fox H-function
with $0<\delta<1$ is always chosen such that it runs from
$c-\mathrm i\infty$ to $c-\mathrm i\infty$ with $0<c<1-\delta$
whereas such straight paths do not exist for 
equation (\ref{eq:fox}) with $\lambda<0$.
Next, we find an extension of equation (\ref{eq:gmlfrd})
for a negative $\lambda=-\xi<0$,
\begin{align}
&E^{-\xi}_{p,b}(\pm z)
=\sum_{k=0}^\mu\binom\xi k\,\frac{(\mp z)^k}{\Gamma(pk+b)}
\\\nonumber
&+\frac{(-1)^{\mu+1}(\delta)_{\mu+1}^+}{\Gamma(1-\delta)}
\RLI0z{1+\xi}\left\lbrace\frac1{z^{1+\xi}}\Biggl[E_{p,b}(\pm z)-
\sum_{k=0}^\mu\frac{(\pm z)^k}{\Gamma(pk+b)}\Biggr]\right\rbrace
\end{align}
where $\mu=\flr\xi$ and $\delta=\xi-\mu$.

\begin{subequations}
Finally we observe additional operational properties that
\begin{align}
&\frac{\rmd^nE^\lambda_{p,b}({-z})}{\rmd z^n}
=(-1)^n(\lambda)_n^+\,E^{\lambda+n}_{p,b+pn}({-z}),
\label{eq:gmld}\\
&\RLI0znE^\lambda_{p,b}({-z})
=\sum_{k=0}^\infty
\frac{(-1)^k(\lambda)_k^+}{\Gamma(b+pk)}\frac{z^{n+k}}{\Gamma(n+k+1)}
\label{eq:gmli}\\\nonumber&\quad
=\frac1{(1-\lambda)_n^+}\left[
E^{\lambda-n}_{p,b-pn}({-z})
-\sum_{k=0}^{n-1}\frac{(n-\lambda)_k^-z^k}{k!\Gamma(b-pn+pk)}\right].
\end{align}
for a non-negative integer $n$.
The last holds given that $(1-\lambda)_n^+\ne0$.
In addtion, using
$(\lambda)_k^+(\lambda+k)=(\lambda)_{k+1}^+=\lambda(\lambda+1)_k^+$,
we also find that
\begin{equation}\label{eq:gmlr}
\frac{\rmd[z^\lambda E^\lambda_{p,b}(-z)]}{\rmd z}
=\sum_{k=0}^\infty
\frac{(-1)^k(\lambda)_{k+1}z^{k+\lambda-1}}{k!\Gamma(b+pk)}
=\zeta z^{\lambda-1}E^{\lambda+1}_{p,b}(-z).
\end{equation}
\end{subequations}

\subsection{Sufficient conditions for the phase-space consistency
of eq.~(\ref{eq:rfun}) with $0<s\le1$}
\label{app:p1}

Now we consider sufficient conditions on the AD to guarantee
the phase-space consistency (Sect.~\ref{sec:suf})
with the radial function given by equation (\ref{eq:rfun})
with $0<s\le1$ and $\mathcal E_0=0$.
In Sect.~\ref{sec:cona}, we have argued that for $\beta_1=\beta_2<1$,
if there exists ${}^\exists\lambda\ge\frac32-\beta_1=\frac32-\beta_2$
such that $\frd0\Psi\lambda P\ge0$ and
$P(0)=\dotsb=P^{(\flr\lambda-1)}=0$, then the df with $\mathcal E_0=0$
recovered from the particular $P(\Psi)$ and $R(x)$ is non-negative
everywhere. This follows from the fact that
$t^{\frac32-\lambda}\phi(t)=t^{\frac32-\lambda-\beta}/\Gamma(1-\beta)$
is cm if and only if $\lambda\ge\frac32-\beta$. As with
$t^{\frac32-\lambda}\phi(t)
=t^{\frac32-\lambda-\beta_1}E^\zeta_{s,1-\beta_1}(-t^s)$ for $\beta_1<1$,
the discussion in Sect.~\ref{sec:cona} on sufficient 
conditions for constant $\beta$ separable AD can carry over here
essentially verbatim if we can establish the set
of results regarding the complete monotonicity of
the generalized Mittag-Leffler functions of the form of
$t^aE^\zeta_{s,1-\beta_1}(-t^s)$.

We first note that the leading term of
$E^\zeta_{s,1-\beta_1}(-t^s)$ for $t\sim0$ is given by
the positive constant
\[E^\zeta_{s,1-\beta_1}(0)=\frac1{\Gamma(1-\beta_1)}>0\]
for $\beta_1<1$,
which indicates that $t^aE^\zeta_{s,1-\beta_1}(-t^s)$
for $\beta_1<1$, $s>0$ and $a>0$
must be increasing in some interval $(0,c)$ where ${}^\exists c>0$.
On the other hand, equation (\ref{eq:gmlasym}) suggests that
\[\lim_{t\rightarrow\infty}t^{\beta_2-\beta_1}E^\zeta_{s,1-\beta_1}(-t^p)
=\frac1{\Gamma(1-\beta_2)}>0\] are positive finite
for $\beta_2<1$. It follows that, as $t\rightarrow+\infty$,
we have $t^aE^\zeta_{s,1-\beta_1}(-t^p)\rightarrow+\infty$
for $\beta_1,\beta_2<1$, $s>0$, and $a>\beta_2-\beta_1$.
That is to say, if $a>\beta_2-\beta_1$, then
$t^aE^\zeta_{s,1-\beta_1}(-t^p)$ with for $\beta_1,\beta_2<1$
and $s>0$ must be increasing in some non-empty subintervals of
$(c,\infty)$ where ${}^\exists c\ge0$. Together,
these observations imply that $t^aE^\zeta_{s,1-\beta_1}(-t^s)$
for $\beta_1<1$, $\beta_2\le1$, and $s>0$ \emph{cannot} be cm
if $a>\min(0,\beta_2-\beta_1)$.
Although it is tempting to hypothesize
by analogy to the constant $\beta$ case such
that $t^aE^\zeta_{s,1-\beta_1}(-t^s)$ for $\beta_1<1$,
$\beta_2\le1$, and $0<s\le1$ are cm for $a\le\min(0,\beta_2-\beta_1)$,
we have only been able to prove this under a restriction that
$\beta_2\ge\beta_1$ or $\beta_2\le1-s$ while for $1-s<\beta_2<\beta_1<1$
we only manage to find a more restrictive condition $a\le-s$
(n.b., $-s<-s+1-\beta_1<\beta_2-\beta_1<0$)
for the complete monotonicity of $t^aE^\zeta_{s,1-\beta_1}(-t^s)$.

We next apply the similar discussion as in Sect.~\ref{sec:cona}
and find that for $R(x)$ given by equation (\ref{eq:rfun})
with $\beta_1<1$, $\beta_2\le1$, and $0<s\le1$,
if there exists
${}^\exists\lambda\ge\frac32-\min(\beta_1,\beta_2)$
such that equations (\ref{eq:cons2}) and (\ref{eq:cons1a})
hold for $P(\Psi)$, then the resulting AD,
$P(\Psi)R(r^2)$ guarantees the existence of a non-negative df,
unless $1-s<\beta_2<\beta_1<1$. For $(s,\beta_2)=(1,1)$, which results in
$E^{1-\beta_1}_{1,1-\beta_1}(-t)={}_1\tilde F_1(1-\beta_1;1-\beta_1;-t)
=\rme^{-t}{}_1\tilde F_1(0;1-\beta_1;t)=\exp(-t)/\Gamma(1-\beta_1)$,
the condition for a positive integer $\lambda=m+1>\frac32-\beta_1$
(note $\min[\beta_1,\beta_2]=\beta_1<1=\beta_2$) reproduces
that of \citet{CM10}, for the generalized \citeauthor{Cu91} system
to result from a non-negative df.
If $1-s<\beta_2<\beta_1<1$ on the other hand, we at this point
only find a slightly restrictive sufficient condition with
${}^\exists\lambda\ge\frac32-(\beta_1-s)>
\frac32-\beta_2>\frac32-\beta_1>\frac12$
(n.b., $\beta_1-s<1-s<\beta_2<\beta_1<1$).

In the rest of this section, we proceed to prove that
$t^aE^\zeta_{s,1-\beta_1}(-t^s)$ where $a=\min(0,\beta_2-\beta_1)$
and $s\zeta=\beta_2-\beta_1$ for $\beta_1<1$, $\beta_2\le1$, and
$0<s\le1$ (but not $1-s<\beta_2<\beta_1<1$) is cm. First,
if $\beta_1=\beta_2<1$, then $E^0_{s,1-\beta_1}(-t^s)=1/\Gamma(1-\beta_1)$
and so is trivial (see Sect.~\ref{sec:cona}). Next
note, if we can prove that $z^{\min(0,\lambda)}E^\lambda_{p,b}(-z)$
is cm for $b>0$, $b\ge p\lambda$, and $0<p\le1$,
then the desired result follows Corollary \ref{cor:cmp}.
In the following, we prove the complete monotonicity of
$E^\lambda_{p,b}(-z)$ for $0<p\lambda\le b$ and
$z^\lambda E^\lambda_{p,b}(-z)$ for $b>0$ and $\lambda<0$.
The further restriction, $\beta_2\le1-s$
(i.e., $p\lambda+p\le b$) on the latter case meanwhile occurs naturally.
We first introduce a lemma,
\begin{lemma}\label{lem:gmlpos}
If $0<p\le1$, $b>0$, and $b\ge p\lambda$, then
$E^\lambda_{p,b}(-z)\ge0$ is non-negative for all $z>0$.
\end{lemma}
This formalizes the fact that $\mathcal R(w)$
for $0<s\le1$ and $\beta_1,\beta_2\le1$ is a cm function of $w>0$.
In general, for $b>0$,
\begin{multline}\label{eq:gmllt}
\int_0^\infty\!\rmd t\,\rme^{-wt}t^{b-1}E^\lambda_{p,b}({-t^p})
=\sum_{k=0}^\infty\!\frac{(-1)^k(\lambda)_k^+}{k!\Gamma(pk+b)}\!\!
\int_0^\infty\!\!\rmd t\,\rme^{-wt}t^{pk+b-1}
\\=\sum_{k=0}^\infty\frac{(-1)^k(\lambda)_k^+}{k!w^{pk+b}}
=\frac1{w^b}\left(1+\frac1{w^p}\right)^{-\lambda}
\end{multline}
By Corollary \ref{cor:qcm}, this is a cm function of $w>0$ for $0<p\le1$
either if $b\ge0$ and $\lambda\le0$ or if $b-p\lambda\ge0$ and
$\lambda\ge0$. Then from the Bernstein theorem,
if $0<p\le1$, $b>0$, and $b\ge p\lambda$,
then $t^{b-1}E^\lambda_{p,b}({-t^p})\ge0$ for $t>0$ and so
$E^\lambda_{p,b}({-z})\ge0$ for $z>0$. 

The first half of the desired result is now trivial, that is,
\begin{theorem}\label{th:pos}
If $0<p\le1$ and $0<p\lambda\le b$, then $E^\lambda_{p,b}(-z)$ is
a cm function of $z>0$.
\end{theorem}
This follows equation (\ref{eq:gmld}).
Note that if $b\ge p\lambda$, then $b+pn\ge\lambda+pn$ and thus
Lemma~\ref{lem:gmlpos} together with $(\lambda)_n^+>0$ for $\lambda>0$
completes the proof. As noted, Theorem~\ref{th:pos} implies
\begin{corollary}
For $\mathcal E_0=0$ and $R(x)$ given by equation (\ref{eq:rfun})
with $0<s\le1$ and $\beta_1<\beta_2\le1$, if there exists
${}^\exists\lambda\ge\frac32-\beta_1$ such that
$\frd0\Psi\lambda P\ge0$ and $P(0)=\dotsb=P^{(\flr\lambda-1)}(0)=0$,
then the df inverted from $P(\Psi)R(r^2)$ is non-negative.
\end{corollary}
This actually extends to $\beta_1\le\beta_2\le1$ (Sect.~\ref{sec:cona}).
Also note that if $P(0)=\dotsb=P^{(\flr{\frac12-\beta})}(0)=0$, then
$\frd0\Psi{\frac32-\beta_1}P\ge0$ is the necessary \emph{and}
sufficient condition for the phase-space consistency
given $\mathcal E_0=0$ and $R(x)$ with $0<s\le1$ and
$\beta_1\le\beta_2\le1$.

For the second half, we first find
\begin{theorem}\label{th:nneg}
If $0<p\le1$, $b>0$, and $\xi\ge0$, then $s^{-\clg\xi}E^{-\xi}_{p,b}(-z)$
is a cm function of $z>0$.
\end{theorem}
\begin{corollary}
For $\mathcal E_0=0$ and $R(x)$ in equation (\ref{eq:rfun})
with $0<s\le1$ and $\beta_2\le\beta_1<1$, if there exists
${}^\exists\lambda\ge\frac32-\beta_1+sn$ where
$n=\clg{(\beta_1-\beta_2)/s}$
such that $\frd0\Psi\lambda P\ge0$
and $P(0)=\dotsb=P^{(\flr\lambda-1)}(0)=0$,
then the df inverted from $P(\Psi)R(r^2)$ is non-negative.
\end{corollary}
If $\xi=\mu$ is a non-negative integer, this is trivial since
\begin{equation}\label{eq:gmln}
z^{-\mu}E^{-\mu}_{p,b}(-z)
=\sum_{k=0}^\mu\binom\mu k\,\frac{z^{-(\mu-k)}}{\Gamma(b+pk)}.
\end{equation}
with every coefficient being positive.
Next, equation (\ref{eq:gmld}) for $\zeta=-\xi\le0$ and $n=\clg\xi$
results in 
\begin{subequations}
\begin{equation}
\frac{\rmd^{\clg\xi}E^{-\xi}_{p,b}({-z})}{\rmd z^{\clg\xi}}=
(1-\epsilon)_{\clg\xi}^+\,E^\epsilon_{p,b+p\clg\xi}({-z}),
\end{equation}
where $0\le\epsilon=\clg\xi-\xi<1$.
Now equation (\ref{eq:gmli}) indicates that
\begin{equation}
(1-\epsilon)_{\clg\xi}^+\RLI0z{\clg\xi}
E^\epsilon_{p,b+p\clg\xi}({-z})
=E^{-\xi}_{p,b}({-z})
-\sum_{k=0}^{\clg\xi-1}\binom\xi k\,
\frac{z^k}{\Gamma(b+pk)},
\end{equation}
which is consistent with equation (\ref{eq:difcomb}).
If $\xi>0$, 
this reduces to (note then that $\clg\xi\ge1$)
\begin{multline}
z^{-\clg\xi}E^{-\xi}_{p,b}({-z})
=\sum_{k=0}^{\clg\xi-1}\binom\xi k\,\frac{s^{-(\clg\xi-k)}}{\Gamma(b+pk)}.
\\+\frac{(1-\epsilon)_{\clg\xi}^+}{(\clg\xi-1)!}
\int_0^1\!\rmd u\,(1-u)^{\clg\xi-1}
E^\epsilon_{p,b+p\clg\xi}({-zu}).
\end{multline}
If $\epsilon=0$ (i.e., $\clg\xi=\xi$), then $(1-\epsilon)_{\clg\xi}^+=\xi!$
and $E^0_{p,b+p\xi}=1/\Gamma(b+p\xi)>0$, and so
this is just equation (\ref{eq:gmln}). In general, this implies
that $z^{-\clg\xi}E^{-\xi}_{p,b}({-z})$
with $0<p\le1$, $b>0$, and $\xi>0$ is cm since
\begin{equation}
\frac{\rmd^n}{\rmd s^n}\!\int_0^1\!\rmd u\,
(1-u)^kf(zu)
=\int_0^1\!\rmd u\,
(1-u)^ku^nf^{(n)}(zu),
\end{equation}
\end{subequations}
while Theorem~\ref{th:pos} indicates that
$f(z)=E^\epsilon_{p,b+p\clg\xi}({-z})$
is cm given $b+p\clg\xi-p\epsilon=b+p\xi>0$.
Finally, we are able to prove
\begin{theorem}
If $0<p\le1$, $\xi>0$, $b>0$, and $b\ge p(1-\xi)$, then
$z^{-\xi}E^{-\xi}_{p,b}(-z)$ is a cm function of $s>0$.
\end{theorem}
\begin{corollary}
For $\mathcal E_0=0$ and $R(x)$ given by equation (\ref{eq:rfun})
with $0<s\le1$, $\beta_2<\beta_1<1$, and $\beta_2\le1-s$,
if there exists ${}^\exists\lambda\ge\frac32-\beta_2$ such that
$\frd0\Psi\lambda P\ge0$ and $P(0)=\dotsb=P^{(\flr\lambda-1)}(0)=0$,
then the df inverted from $P(\Psi)R(r^2)$ is non-negative.
\end{corollary}
If $\xi$ is a positive integer, this is the same as
Theorem~\ref{th:nneg}. For general cases,
we note equation (\ref{eq:gmlr}) results in
\begin{equation}
\frac{\rmd[z^{-\xi}E^{-\xi}_{p,b}(-z)]}{\rmd z}
=-\frac{\xi E^{1-\xi}_{p,b}(-z)}{z^{\xi+1}}
=-\frac{\xi z^{-\clg{\xi-1}}E^{-(\xi-1)}_{p,b}(-z)}
{z^{2-\epsilon}}
\end{equation}
where $\clg{\xi-1}=\clg\xi-1$ and $0\le\epsilon=\clg\xi-\xi<1$.
Theorem~\ref{th:nneg} indicates that
if $0<p\le1$, $b>0$, and $\xi\ge1$, then
$z^{\flr{1-\xi}}E^{1-\xi}_{p,b}(-z)$ and subsequently
$z^{-(\xi+1)}E^{1-\xi}_{p,b}(-z)$ are cm.
Theorem~\ref{th:pos} on the other hand suggests that
if $0<p\le1$, $\xi<1$, and $b\ge p(1-\xi)$, then
$E^{1-\xi}_{p,b}(-z)$ is cm. Hence,
if $0<p\le1$, $\xi>0$, $b>0$, and $b\ge p(1-\xi)$,
the derivative of $z^{-\xi}E^{-\xi}_{p,b}(-z)$ is given by
a cm function multiplied by a negative constant. It follows
Lemma \ref{lem:cm}-\textit2
that $z^{-\xi}E^{-\xi}_{p,b}(-z)$ for $0<p\le1$, $\xi>0$,
$b>0$ and $b\ge p(1-\xi)$ is a cm function of $z>0$.

\section{Summary}
\label{sec:sum}

We have shown that the fractional calculus operations
(eqs.~\ref{eq:mint}, \ref{eq:frd}, and \ref{eq:mintn})
applied to the bivariate augmented density
(eq.~\ref{eq:ad}) result in a set of the integral transformations
of the two-integral distribution function
(eqs.~\ref{eq:dpirn2}, \ref{eq:dripn2}, and \ref{eq:last}).
Equation (\ref{eq:last}) with $\lambda+\xi+\frac12=0$
indicates that the set of fractional calculus operations
on the augmented density $\Nu(\Psi,r^2)$ listed in
equation (\ref{eq:msad}) provides with the complete
moment sequence of the distribution function along
$\mathcal K(\mathcal E,L^2;\Psi,r^2)=0$ as shown in
equation (\ref{eq:moms}). We infer from this that
the augmented density that ensures the non-negativity
of the distribution function may be deduced by analogy
to the classical moment problem in probability theory \citep{vH11}.
We have also found that equation (\ref{eq:last})
for a non-negative integer $\lambda>0$ and $\xi=0$
consists in the complete moment
sequence of the augmented density at a fixed $r$ considered
as a probability density on $\Psi$ -- which is possible because the
augmented density is also non-negative in all accessible $r$ and $\Psi$.
Comparing this sequence to the velocity moments resulting
from the given distribution function (eq.~\ref{eq:dist}),
we deduce that the augmented density
(and subsequently the distribution function) is uniquely specified
given the potential $\Psi(r)$ and the density profile $\nu(r)$
once the infinite set of the radial velocity moments in every order
(equivalently the complete radial velocity distribution)
as a function of the radius are available \cite[cf.,][]{DM92}.

Given $\lambda+\xi+\frac12\ge0$, all the integrands
in the right-hand sides of equation (\ref{eq:last}) are non-negative
because the distribution function $\mathcal F(\mathcal E,L^2)$ must
be non-negative in the whole accessible subspace volume $\mathcal T$.
This non-negativity implies that
it is necessary for the integro-differential operations on
the augmented density $\Nu(\Psi,r^2)$ given in
the left-hand side of equation (\ref{eq:last}) to be also non-negative,
\emph{provided that the integrals involved in their definitions
are all convergent}. This introduces the set of necessary
conditions on the augmented density for the non-negativity of
the distribution function.
If the augmented density is multiplicatively separable into
functions of the potential and the radius dependencies like
equation (\ref{eq:sep}), this results in the condition
stated by \citet{An11}, that is, equation (\ref{eq:main1})
for the radius part of the augmented density. We have also
discovered a few equivalent statements of this condition, notably
equation (\ref{eq:main4}) and the function $\mathcal R(w)$
defined in equation (\ref{eq:rlap}) being completely monotonic
and so on. The same argument for the potential part of a separable
augmented density on the other hand recovers the conditions
derived by \citet{vHBD11} and \citet{An11a}. They are further
generalized with fractional calculus to indicate that:
$\frd{\mathcal E_0}\Psi\mu P\ge0$ for
all accessible $\Psi$ is necessary if $\mu\le\frac12$ or
there exists ${}^\exists\lambda\ge\mu-\frac12$ such that
$\RLI0{r^2}\lambda[r^{-2\lambda}R(r^2)]$ is well-defined
or ${}^\exists\beta\le\frac32-\mu$ such that
$\lim_{r^2\rightarrow0^+}r^{2\beta}R(r^2)$ is non-zero and finite.

With separable augmented densities, the distribution function
may be inverted from the augmented density by means of
the inverse Laplace transform as in equation (\ref{eq:lapdf}).
The non-negativity of the distribution
function corresponding to a separable augmented density
is guaranteed if the Laplace transformation of the
distribution function given in equation (\ref{eq:elap})
is a complete monotonic function of $s>0$ for any $L^2\ge0$.
We have shown from this that the set of joint conditions
composed of equation (\ref{eq:rcon0}) with all non-negative integer pairs
$n$ and $k$ for the radius part $R(r^2)$ of the augemented density
and equations (\ref{eq:cons2}) and (\ref{eq:cons1a})
for the potential part $P(\Psi)$ of the same is sufficient
to imply the non-negativity of the corresponding distribution
function. This last set of sufficient conditions
is equivalent to that of \citet{vH11}, which was derived
from the argument following the application of the Hausdorff moment
problem.

\small
\acknowledgments\noindent
This manuscript is basically an extended version
of \citet{An12}.

\newpage\normalsize
\section*{appendix}
\setcounter{section}0
\numberwithin{equation}{section}
\renewcommand\thesection{\Alph{section}}
\renewcommand\theequation{\thesection\arabic{equation}}

\section{proper derivations of \lowercase{$\phi(t)$} in
E\lowercase{q.~(\ref{infs})}}\label{sec:app}

Let us define with $R(x)$ given by equation (\ref{eq:rfun}) so that
\begin{subequations}
\begin{gather}
\alpha_n\equiv\frac{R_{(n)}}R\,;\qquad
\tau_n\equiv(1+x^s)^n\alpha_n,
\\
u\equiv\frac{x^s}{1+x^s}\,;\qquad
y\equiv x^s.
\end{gather}
\end{subequations}
Using equation (\ref{eq:lem}), we find
the recursion formula for $\alpha_n(u)$,
\begin{subequations}
\begin{align}
\alpha_{n+1}
&=\frac1{x^nR}\frac{\rmd(x^{n+1}R\alpha_n)}{\rmd x}
=\frac{\rmd\log(x^{n+1}R)}{\rmd\log x}\alpha_n
+u\frac{\rmd\log u}{\rmd\log x}\frac{\rmd\alpha_n}{\rmd u}
\nonumber\\&=\bigl[n+1-\beta_1+(\beta_1-\beta_2)u\bigr]\,\alpha_n
+su(1-u)\frac{\rmd\alpha_n}{\rmd u}.
\label{eq:arec}\end{align}
Given that
\begin{equation}
\frac{\rmd\alpha_n}{\rmd u}
=\frac{\rmd y}{\rmd u}\frac\rmd{\rmd y}\biggl[\frac{\tau_n}{(1+y)^n}\biggr]
=\frac1{(1+y)^{n-2}}\frac{\rmd\tau_n}{\rmd y}
-\frac{n\tau_n}{(1+y)^{n-1}},
\end{equation}
the recursion formula for $\tau_n(y)$,
\begin{align}
\tau&_{n+1}=(1+y)^{n+1}\alpha_{n+1}
\\&=\biggl[n+1-\beta_1+\frac{(\beta_1-\beta_2)y}{1+y}\biggr]\,
(1+y)^{n+1}\alpha_n
+sy(1+y)^{n-1}\frac{\rmd\alpha_n}{\rmd u}
\nonumber\\&=\bigl[n+1-\beta_1+(n+1-\beta_2-pn)y\bigr]\,\tau_n
+sy(1+y)\frac{\rmd\tau_n}{\rmd y},
\nonumber\end{align}
\end{subequations}
also follows. Both recursion formulae imply that
$\alpha_n(u)$ and $\tau_n(y)$ are an (at most) $n$-th order polynomial
of their respective arguments, $u$ and $y$
(note $\alpha_0=\tau_0=1$ by definition).
Subsequently, if we let
\begin{subequations}
\begin{gather}
\alpha_n=\textstyle{\sum_{k=0}^n\tilde a_{n,k}u^k}
\,;\qquad
\tau_n=\textstyle{\sum_{k=0}^n\tilde t_{n,k}y^k},
\intertext{then}
R_{(n)}(x)=\sum\nolimits_{k=0}^n
\frac{\tilde a_{n,k}x^{sk-\beta_1}}{(1+x^s)^{\zeta+k}}
=\frac{\sum_{k=0}^n\tilde t_{n,k}x^{sk-\beta_1}}{(1+x^s)^{\zeta+n}}.
\label{c2a}\end{gather}
\end{subequations}
In addition, given that $\tau_n=(1+y)^n\alpha_n$,
the binomial expansion and the subsequent rearrangement of the double sum
\begin{subequations}
\begin{align}
\tau_n=(1+y)^n\alpha_n
&=\sum_{k=0}^n\tilde a_{n,k}y^k(1+y)^{n-k}
=\sum_{k=0}^n\sum_{j=0}^{n-k}\binom{n-k}j\,\tilde a_{n,k}y^{k+j}
\nonumber\\&
=\sum_{m=0}^n\sum_{k=0}^m\frac{(n-k)!}{(m-k)!(n-m)!}\tilde a_{n,k}y^m,
\end{align}
leads to the relation between the two sets of coefficients,
\begin{equation}\label{eq:ct}
\tilde t_{n,m}
=\frac1{(n-m)!}\sum\nolimits_{k=0}^m\frac{(n-k)!}{(m-k)!}\tilde a_{n,k}.
\end{equation}
We note that if we define the associated coefficient sets,
\begin{equation}
\tilde t_{n,k}=\binom nk\,t_{n,k}
\,;\qquad
\tilde a_{n,k}=(-1)^k\binom nk\,a_{n,k}
\end{equation}
the relation reduces to the standard binomial transform,
\begin{equation}
t_{n,k}=\sum_{m=0}^k(-1)^m\binom km\,a_{n,m}
\quad\Longleftrightarrow\quad
a_{n,k}=\sum_{m=0}^k(-1)^m\binom km\,t_{n,m},
\end{equation}
\end{subequations}
which is known to be involutionary.\newpage

The expression for the coefficients is found
using the binomial series expansion of equation (\ref{c2a})
for $0\le x<1$,
\begin{subequations}
\begin{equation}\label{c2}\begin{split}
R_{(n)}(x)&=\sum_{k=0}^n\tilde a_{n,k}
\sum_{j=0}^\infty\frac{(-1)^j(\zeta+k)_j}{j!}x^{s(k+j)-\beta_1}
\\&=\sum_{m=0}^\infty\sum_{k=0}^m
\tilde a_{n,k}\frac{(-1)^{m-k}(\zeta+k)_{m-k}}{(m-k)!}x^{sm-\beta_1}
\end{split}\end{equation}
where we have used $\tilde a_{n,k}=0$ if $k>n$.
All the Pochhammer symbols without any directional specification
hereafter are interpreted to represent
the rising product, i.e., $(a)_n=(a)_n^+$.
Matching the coefficients for the same power of $x$ in
equations (\ref{c2}) and (\ref{c3}) leads to
[n.b., $(\zeta)_m(\zeta+m)_{k-m}=(\zeta)_k$]
\begin{equation}
(1-\beta_1+sk)_n
=\sum\nolimits_{m=0}^k\frac{(-1)^mk!}{(k-m)!}
\frac{\tilde a_{n,m}}{(\zeta)_m}.
\end{equation}
Although its derivation assumed $0\le x<1$, this is valid
regardless. The right-hand side is in the form of
the binomial transformation and thus by its involutionary inversion
\begin{equation}\label{c4}\begin{split}
\tilde a_{n,m}
&=(\zeta)_m\sum\nolimits_{k=0}^m\frac{(-1)^k}{k!(m-k)!}\,(1-\beta_1+sk)_n
\\&=\frac{(-1)^m(\zeta)_m}{m!}\,\Delta_x^m(1-\beta_1+sx)_n\bigr\rvert_{x=0}.
\end{split}\end{equation}
\end{subequations}
That is, $\tilde a_{n,k}$ is the $k$-th order forward
finite difference of $(1-\beta_1+sx)_n=\prod_{j=1}^n(j-\beta_1+sx)$
at $x=0$. Since $(1-\beta_1+sx)_n$ is an $n$-th order polynomial
of $x$, we have $\Delta_x^k(1-\beta_1+sx)_n\rvert_{x=0}=0$ if $k>n$.
The formula for $\tilde t_{n,k}$ is found from equation (\ref{c4}),
\begin{subequations}
\begin{align}
\tilde t_{n,k}&=\sum\nolimits_{m=0}^k\frac{(-1)^m}{m!(k-m)!}
(\zeta)_m^+(n+\zeta)_{k-m}^-(1-\beta_1+sm)_n^+
\nonumber\\&=(\zeta)_{n+1}\sum\nolimits_{m=0}^k\frac{(-1)^m}{m!(k-m)!}
\frac{(1-\beta_1+sm)_n}{(\zeta+m)_{1+n-k}},
\end{align}
using equation (\ref{eq:ct}) and the Chu\footnote{Zh\=u Sh\`iji\'e
(1270-1330)}--Vandermonde\footnote{Alexandre-Th\'eophile Vandermonde
(1735-1796)} identity
\begin{equation}
\textstyle\sum_{k=0}^n\binom nk\,(s)_k(t)_{n-k}=(s+t)_n,
\end{equation}
or equivalently the Gauss\footnote{Johann Carl Friedrich
Gauss (1777-1855)} hypergeometric identity
\begin{equation}
\sum\nolimits_{k=0}^n(-1)^k\binom nk\,\frac{(b)_k}{(c)_k}
=\sum\nolimits_{k=0}^n\frac{(-n)_k(b)_k}{k!(c)_k}
=\frac{(c-b)_n}{(c)_n}.
\label{ghid}\end{equation}
\end{subequations}

For the $s=1$ case, from equations (\ref{c4}), (\ref{ghid}), and
\begin{subequations}
\begin{equation}
(1-\beta_1)_k(1-\beta_1+k)_n
=(1-\beta_1)_n(1-\beta_1+n)_k
\end{equation}
%
we find that ($\beta_1<1$)
\begin{align}
\frac{\tilde a_{n,m}}{(1-\beta_1)_n}
&=\frac{(\beta_2-\beta_1)_m(-n)_m}{m!(1-\beta_1)_m}
=(-1)^m\binom nm\,\frac{(\beta_2-\beta_1)_m}{(1-\beta_1)_m};\\
\frac{t_{n,k}}{(1-\beta_1)_n}
&=\sum_{m=0}^k(-1)^m\binom km\,\frac{(\beta_2-\beta_1)_m}{(1-\beta_1)_m}
=\frac{(1-\beta_2)_k}{(1-\beta_1)_k}.
\end{align}
Here $\zeta=\beta_2-\beta_1$ since $s=1$.
This is notable as it indicates that
$\tilde t_{n,k}\ge0$, and with
$(1-\beta_1)_k(1-\beta_1+k)_{n-k}=(1-\beta_1)_n$ that
\begin{gather}
\tau_n(y)
=\sum_{k=0}^n\binom nk\,(1-\beta_1+k)_{n-k}(1-\beta_2)_ky^k
\ge0
\\
R_{(n)}(x)
=\frac1{(1+x)^{\beta_2-\beta_1+n}}
\sum_{k=0}^n\binom nk\,(1-\beta_1+k)_{n-k}(1-\beta_2)_kx^{k-\beta_1}
\end{gather}\end{subequations}
for any non-negative integer $n$ and all $x=y>0$.

Next, we consider the Mellin\footnote{Robert Hjalmar Mellin (1854-1933)}
transform for $0<z<\lambda$
\begin{subequations}
\begin{align}
\varphi&(z)
=\underset{y\rightarrow z}{\mathcal M}
\biggl[\frac{\alpha_n}{(1+y)^\lambda}\biggr]
=\int_0^\infty\!\rmd y\,y^{z-1}\frac{\alpha_n}{(1+y)^\lambda}
\\\nonumber&=\int_0^\infty\!\rmd u\,(1-u)^{\lambda-z-1}u^{z-1}\alpha_n
=\sum_{k=0}^n
\frac{\Gamma(z+k)\Gamma(\lambda-z)}{\Gamma(k+\lambda)}\tilde a_{n,k}
\\\nonumber&=\frac{\Gamma(z)\Gamma(\lambda-z)}{\Gamma(\lambda)}\sum_{k=0}^n
\frac{(-1)^k(z)_k(\zeta)_k}{k!(\lambda)_k}\,
\Delta_x^k(1-\beta_1+sx)_n\bigr\rvert_{x=0}.
\end{align}
This simplifies for $\lambda=\zeta$ utilizing the Newton\footnote{Sir
Isaac Newton (1642-1727)} series
\begin{equation}
f(z)=\sum\nolimits_{k=0}^\infty
\frac{(z)_{k}^-}{k!}\Delta_x^kf(x)\bigr\rvert_{x=0}.
\end{equation}
If $f(x)$ is an $n$-th order polynomial, the formula is
exact after the summation up to $k=n$.
Since $\Delta_x^k(1-\beta_1+sx)_n\rvert_{x=0}=0$ for $k>n$
with the $n$-th order polynomial $(1-\beta_1+sx)_n$,
\begin{equation}
(1-\beta_1-sz)_n=\sum\nolimits_{k=0}^n
\frac{(-1)^k(z)_k}{k!}\Delta_x^k(1-\beta_1+sx)_n\bigr\rvert_{x=0}.
\end{equation}
and therefore with $\lambda=\zeta>z>0$,
\begin{equation}
\Gamma(\zeta)\varphi(z)=\Gamma(z)\Gamma(\zeta-z)(1-\beta_1-sz)_n.
\end{equation}
By means of the inverse Mellin transformation, $R_{(n)}=R\alpha_n$ is
then expressible to be a Mellin--Barnes\footnote{Ernest William Barnes
(1874-1953)} type integral
\begin{equation}
R_{(n)}(x)=\frac1{2\pi\mathrm i\,x^{\beta_1}}\!
\int_\mathcal C\frac{\rmd z}{x^{sz}}
\frac{\Gamma(z)\Gamma(\zeta-z)(1-\beta_1-sz)_n}{\Gamma(\zeta)}.
\end{equation}
\end{subequations}
Although this is actually reducible to algebraic functions on $x^s$
as in equation (\ref{c2a}), it is also the \citeauthor{Fx61} H-function
($H^{1,2}_{2,2}$ in particular) and further reduces to
the Me{\ij}er\footnote{Cornelis Simon Me{\ij}er (1904-1974)} G-function
($G^{1,n+1}_{n+1,n+1}$) and the hypergeometric function (${}_{n+1}F_n$),
the last of which would be formally equivalent to equation (\ref{c3}).

%

The function $\phi(t)$ is found from equation (\ref{eq:phit})
\begin{subequations}
\begin{align}
\phi(t)&=\frac1{2\pi\mathrm i\,t^{\beta_1}}\!
\int_\Cr\!\frac{\rmd z}{t^{sz}}
\frac{\Gamma(z)\Gamma(\zeta-z)}{\Gamma(\zeta)}
\lim_{n\rightarrow\infty}\frac{(1-\beta_1-sz)_n\,n^{\beta_1+sz}}{n!}
\nonumber\\&=\frac1{2\pi\mathrm i\,t^{\beta_1}}\!
\int_\Cr\!\frac{\rmd z}{t^{sz}}
\frac{\Gamma(z)\Gamma(\zeta-z)}{\Gamma(\zeta)\Gamma(1-\beta_1-sz)}
\nonumber\\&=\frac1{t^{\beta_1}\Gamma(\zeta)}\,
H^{1,1}_{1,2}\!\left\lgroup t^s\,\vrule\begin{array}c
\{1-\zeta,1\}\\\{0,1\},\{\beta_1,s\}
\end{array}\right\rgroup.
\end{align}
%
where we have used equation (\ref{c5}).
If $0<\zeta\le(1-\beta_1)/s$
(n.b., $\zeta=(\beta_2-\beta_1)/s$ and $\beta_2\le1$),
the convergent integration path $\Cr$ 
may be chosen such that $z=c-i\infty$ to $z=c+i\infty$ with $0<c<\zeta$.
Provided that $\zeta$ is neither zero nor a negative integer,
this is still valid but the integration path
should rather be chosen to separate the poles of $\Gamma(z)$
from those of $\Gamma(\zeta-z)$.

In fact, $\phi(t)$ may alternatively be found in terms of
an infinite series for any $\zeta$. In particular,
equation (\ref{eq:phit}) after inserting equation (\ref{c4})
into equation (\ref{c2a}) results in
\begin{align}
R_{(n)}&=\sum_{k=0}^n\sum_{q=0}^k
\frac{(-1)^q(\zeta)_k}{q!(k-q)!}(1-\beta_1+sq)_n
\frac{x^{sk-\beta_1}}{(1+x^s)^{\zeta+k}}
\\\phi(t)&=\lim_{n\rightarrow\infty}\sum_{k=0}^n\sum_{q=0}^k
\frac{(-1)^q(\zeta)_k}{q!(k-q)!}\frac{(1-\beta_1+sq)_n}{n!}
\frac{x^{sk-\beta_1}}{(1+x^s)^{\zeta+k}}\biggr\rvert_{x=t/n}
\nonumber\\&
=\sum_{k=0}^\infty\sum_{q=0}^k
\frac{(-1)^q(\zeta)_kt^{sk-\beta_1}}{q!(k-q)!}
\lim_{n\rightarrow\infty}\frac{(1-\beta_1+sq)_n}
{n!n^{sk-\beta_1}(1+t^s/n^s)^{\zeta+k}}
\nonumber\\&
=\sum\nolimits_{k=0}^\infty
\frac{(-1)^k(\zeta)_k}{k!\Gamma(1-\beta_1+sk)}t^{sk-\beta_1}
\nonumber\\&
=\frac1{t^{\beta_1}\Gamma(\zeta)}\,{}_1\Psi_1\!\left[\begin{array}c
(\zeta,1);\\(1-\beta_1,s);\end{array}{-t^s}\right],
\end{align}
where we have used equation (\ref{c5}) and
\begin{equation}
\lim_{n\rightarrow\infty}\frac1{n^a}
=\begin{cases}
1&(a=0)\\0&(a>0)
\end{cases}.
\end{equation}
\end{subequations}

For $s=1$, we have
$\phi(t)=t^{-\beta_1}{}_1\tilde F_1(\beta_2-\beta_1;1-\beta_1;-t)$
where ${}_1\tilde F_1(a;b;x)={}_1F_1(a;b;x)/\Gamma(b)$ is the
regularized hypergeometric function. The non-negativity of $\phi(t)\ge0$
for $t>0$ and $s=1$ is explicitly shown by the Kummer
hypergeometric transformation resulting in
$\phi(t)=t^{-\beta_1}\rme^{-t}{}_1\tilde F_1(1-\beta_2;1-\beta_1;t)$.
That is to say, from equation (\ref{infs}) in general
\begin{subequations}
\begin{equation}\begin{split}
t^{\beta_1}\exp(t^s)\,\phi(t)&=\sum_{k,m=0}^\infty
\frac{(-1)^k(\zeta)_k}{k!m!\Gamma(1-\beta_1+sk)}t^{s(k+m)}
\\&=\sum_{n=0}^\infty\frac{t^{sn}}{n!}
\sum_{k=0}^n(-1)^k\binom nk\,\frac{(\zeta)_k}{\Gamma(1-\beta_1+sk)}.
\end{split}\end{equation}
The inner sum for $s=1$ simplifies, from equation (\ref{ghid}), to
%
\begin{gather}
\sum_{k=0}^n(-1)^k\binom nk\,
\frac{(\beta_2-\beta_1)_k}{\Gamma(1-\beta_1+k)}
=\frac{(1-\beta_2)_n}{\Gamma(1-\beta_1+n)},
\intertext{and therefore for $t>0$}
\phi(t)
=\frac{\rme^{-t}}{t^{\beta_1}}\sum\nolimits_{n=0}^\infty
\frac{(1-\beta_2)_n}{n!\Gamma(1-\beta_1+n)}\,t^n\ge0,
\end{gather}\end{subequations}
which is non-negative for all pairs $(\beta_1,\beta_2)$
given that $\beta_1,\beta_2\le1$ as every coefficient
of the series is then non-negative as well.

\section{The $\beta_1=1$ cases}\label{app:b1}

\subsection{A proof of eq.~(\ref{eq:dlim})}

First, we note a trivial result,
\begin{lemma}\label{lem:b1}
For $c,\lambda>0$,
\[\lambda\!\int_0^cx^{\lambda-1}\,\rmd x=c^\lambda\,;\qquad
\lim_{\lambda\rightarrow0^+}c^\lambda=1.\]
\end{lemma}
Next, it follows that
\begin{theorem}
for $F(x)=f(x)-\ell$ where $\ell=\lim_{x\rightarrow0^+}f(x)$,
\[\lim_{\lambda\rightarrow0^+}\lambda\!
\int_0^c\!x^{\lambda-1}\abs{F(x)}\,\rmd x=0
\qquad(c>0)\]
\smallskip\\{\it proof.}
First by the definition of $\lim_{x\rightarrow0^+}f(x)$, we find that
for any ${}^\forall\epsilon>0$, there exists
${}^\exists\delta>0$ such that, if $0<x<\delta$,
then $\abs{f(x)-\ell}=\abs{F(x)}<\epsilon$. Now if $0<c\le\delta$, then
for any $\lambda>0$
\[0\le\int_0^c\!x^{\lambda-1}\abs{F(x)}\,\rmd x
<\epsilon\int_0^c\!x^{\lambda-1}\rmd x.\]
If $c>\delta>0$ on the other hand,
\begin{multline*}
0\le\int_0^c\!x^{\lambda-1}\abs{F(x)}\,\rmd x
=\int_0^\delta\!x^{\lambda-1}\abs{F(x)}\,\rmd x
+\int_\delta^c\!x^{\lambda-1}\abs{F(x)}\,\rmd x
\\<\epsilon\int_0^\delta\!x^{\lambda-1}\rmd x
+\sup_{(\delta,c)}[\abs{F(x)}]\int_\delta^c\!x^{\lambda-1}\,\rmd x.
\end{multline*}
Note here that $\int_\delta^c\!x^{\lambda-1}\,\rmd x$ is finite.
Consequently, provided that $f(x)$ is bounded in $(0,c)$,
we find from both cases that
\[0\le\lim_{\lambda\rightarrow0^+}\lambda\!
\int_0^c\!x^{\lambda-1}\abs{F(x)}\,\rmd x<\epsilon\]
where $c>0$ and we have used Lemma \ref{lem:b1}. {\sc q.e.d.}
\end{theorem}
It immediately follows that
\begin{corollary}
for $F(x)=f(x)-\ell$ where $\ell=\lim_{x\rightarrow0^+}f(x)$,
\[\begin{split}
\lim_{\lambda\rightarrow0^+}&\lambda\!
\int_0^c\!x^{\lambda-1}F(x)\,\rmd x=0
\qquad(c>0),\\\lim_{\lambda\rightarrow0^+}&\lambda\!
\int_0^c\!x^{\lambda-1}f(x)\,\rmd x=\ell
\qquad(c>0).\end{split}\]
\end{corollary}
Equation (\ref{eq:dlim}) trivially follows this with the change
of integration variable $x=t-a$. Formally this is interpreted to be
\begin{equation}\label{eq:delta}
\lim_{\lambda\rightarrow0^+}\lambda x^{\lambda-1}=\deltaup(x)
\,;\qquad
\lim_{a\rightarrow1^-}\frac1{x^a\Gamma(1-a)}=\deltaup(x)
\end{equation}
where $\deltaup(x)$ is the Dirac\footnote{Paul Adrien Maurice Dirac
(1902-1984)} delta, provided that $f(x)$ is right-continuous.

\subsection{The $\beta=1$ constant anisotropy model}

Let us consider the df given by
\begin{subequations}
\begin{equation}
\mathcal F(\mathcal E,L^2)=
\frac{f(\mathcal E)\,\deltaup(L^2)}{\sqrt2\pi^\frac32}
\end{equation}
where $f(\mathcal E)$ is an arbitrary function of $\mathcal E$.
This df corresponds to the spherical system entirely built
by radial orbits, that is,
the $\beta=1$ constant anisotropy model.
Given that $\mathcal K(L^2=0)=2(\Psi-\mathcal E)$,
the corresponding AD is found to be
\begin{equation}
\Nu(\Psi,r^2)=\frac1{r^2}\!\sqrt{\frac2\pi}\!
\int_{\mathcal E_0}^\Psi\!
\frac{f(\mathcal E)\,\rmd\mathcal E}{\sqrt{2(\Psi-\mathcal E)}}
=r^{-2}\RLI{\mathcal E_0}\Psi{\frac12}f(\Psi),
\end{equation}
which is separable as in equation (\ref{eq:sep}) with
\begin{equation}
P(\Psi)=\RLI{\mathcal E_0}\Psi{\frac12}f(\Psi)
\,;\qquad
R(x)=x^{-1}.
\end{equation}
The AD is inverted to the df using the fractional derivative,
\begin{equation}
f(\mathcal E)=\frd{\mathcal E_0}{\mathcal E}{\frac12}P(\mathcal E)\ge0,
\end{equation}
\end{subequations}
whose non-negativity is also the necessary \emph{and}
sufficient condition for the phase-space consistency.
Note that this is consistent with the results of Sect.~\ref{sec:cona}
applicable for $\beta\le1$ as is $R(x)$ here the natural limit of
the constant anisotropy model in equation (\ref{eq:cbeta}) to $\beta=1$.

Furthermore, we find for $\lambda=n+\delta>0$ and $n=\flr\lambda$ that
\begin{subequations}\begin{gather}
\RLI0x\lambda x^{-1-\lambda}=\frac1{\Gamma(\lambda)}
\int_0^x\!\frac{(x-y)^{\lambda-1}\,\rmd y}{y^{\lambda+1}}
\rightarrow\infty,
\\
\RLI0x{1-\delta}x^{\lambda-1}=\frac1{\Gamma(1-\delta)}
\int_0^x\!\frac{y^{\lambda-1}\,\rmd y}{(x-y)^\delta}
=\frac{x^n\Gamma(\lambda)}{n!};
\\
\frd0x\lambda x^{\lambda-1}
=\frac{\rmd^{n+1}}{\rmd x^{n+1}}\RLI0x{1-\delta}x^{\lambda-1}
=\frac{\Gamma(\lambda)}{n!}\frac{\rmd^{n+1}x^n}{\rmd x^{n+1}}=0,
\end{gather}\end{subequations}
while $\RLI0x0x^{-1}=\frd0x0x^{-1}=x^{-1}$. Hence,
$R(x)=x^{-1}$ satisfies the necessary condition
in equation (\ref{eq:main1}).
Moreover, equations (\ref{eq:dpirn2}), (\ref{eq:dripn2}),
and (\ref{eq:last}) still hold with non-trivial cases indicating
\begin{equation}
\frd{\mathcal E_0}\Psi\mu P
=\RLI{\mathcal E_0}\Psi{\frac12-\mu}f(\Psi),
\end{equation}
whose non-negativity for ${}^\forall\mu\le\frac12$ is
the same necessary condition for $P(\Psi)$ discussed
in Sect.~\ref{sec:potn}.

From $R(x)=x^{-1}$, we find that $\mathcal R(w)=1$ and
its inverse Laplace transformation at least formally is
given by $\phi(t)=\deltaup(t)$. Although equation (\ref{eq:rcon})
strictly is then trivial as $\deltaup(t)=0$ for $t>0$, this
interpretation of equation (\ref{eq:rcon}) seems improper
considering that the Dirac delta is not differentiable at $t=0$.
Equation (\ref{eq:rcon0}) on the other hand reduces to
$x^{\frac12-\lambda}$ being cm since $R_{(0)}(x)=R(x)=x^{-1}$
and $R_{(n)}(x)=0$ for any positive integer $n$. The sufficient
condition following this, that is, equations (\ref{eq:cons2})
and (\ref{eq:cons1a}) for ${}^\exists\lambda\ge\frac12$ is in fact
a proper one, as is the natural limiting case of the constant
anisotropy model for $\beta=1$. It appears that for $R\sim x^{-1}$
as $x\sim0$ (and $\lim_{w\rightarrow\infty}\mathcal R$ being
nonzero finite), we may consider $\phi(t)\sim t^{-1}$
as $t\sim0$ for the purpose of applying equation (\ref{eq:rcon}).

\subsection{Equation (\ref{eq:rfun}) with $\beta_1=1$}

The discussion in Sect.~\ref{sec:nec} on necessary conditions
is valid inclusively for $\beta_1\le1$. That is,
equation (\ref{eq:rfun}) with $\beta_1=1$ still requires to satisfy
equation (\ref{eq:main1}) -- if $0<p\le1$, this is automatically met
-- in order for the df to be non-negative
whereas the potential dependent part is restricted to be
$\frd{\mathcal E_0}\Psi{\frac12}P\ge0$ for the phase-space
consistency.

The complication arises however for $\beta_1=1$
in regards to sufficient conditions discussed
in the preceding section. The main difficulty is due to the fact that
$\lim_{x\rightarrow0}xR(x)=\lim_{w\rightarrow\infty}\mathcal R(w)=1$
is non-zero.
Whilst this would indicate $\phi\sim t^{-1}$ for $t\sim0$, 
%
%
the particular behavior is incompatible with the convergence of the
integral. The formal solution follows adopting equation (\ref{eq:delta}).
In addition, the limit of equation (\ref{eq:main4}) with $R=x^{-1}$
is identically zero for any $x>0$ and so the function $\phi(t)$
defined via the formal limit of equation (\ref{eq:phit})
with $R(x)$ in equation (\ref{eq:rfun}) takes the same value
as that with ``$R(x)-x^{-1}$'' for all $t>0$ (that is to say,
the Post--Widder formula is technically valid). In other words, 
the function $\phi(t)$ derived in equation \ref{infs}
with $\beta_1=1$ is in fact the inverse Laplace
transform of ``$\mathcal R(w)-1$'' and the `true' inverse transformation
of $\mathcal R(w)$ with $\beta_1=1$ is given by $\phi(t)+\delta(t)$.
For example, since $1/\Gamma(0)=0$, the $k=0$ term in the power series
defining the generalized Mittag-Leffler function $E^\zeta_{p,0}$
does not contribute. Hence, 
equation (\ref{eq:gmllt}) can in fact be well-defined
for the $b=0$ case too. In particular,
\begin{equation*}
\int_0^\infty\!\rmd t\,\rme^{-wt}t^{-1}E^\zeta_{p,0}({-t^p})
=\sum\nolimits_{k=1}^\infty\frac{(-1)^k(\zeta)_k}{k!w^{pk}}
=\left(1+\frac1{w^p}\right)^{-\zeta}-1.
\end{equation*}
Since $(1+w^{-p})^{-\zeta}\ge1$ for $w>0$ and $\zeta\le0$,
it follows that, if $0<p\le1$ and $\zeta\le0$,
this is also cm and $E^\zeta_{p,0}(-z)\ge0$ for $p>0$.
Given that $\mathcal L_{t\rightarrow w}[\deltaup(t)]=1$, we also find
from this that
%
\begin{equation}
\laplace tw\bigl[\delta(t)+t^{-1}E^{-\xi}_{p,0}({-t^p})\bigr]
=(1+w^{-p})^\xi.
\end{equation}

For the specific discussion concerning sufficient conditions for the
non-negativity of the df, we basically consider
\begin{equation*}
P(\Psi)R(r^2)=P(\Psi)R_0(r^2)+r^{-2}P(\Psi)
\end{equation*}
where $R_0(x)=R(x)-x^{-1}$. The corresponding df ($\mathcal E_0=0$) would be
\begin{equation*}
\mathcal F(\mathcal E,L^2)=
\ilaplace s{\mathcal E}\biggl[\frac{s^\frac32\mathcal P(s)}{(2\pi)^\frac32}
\phi\Bigl(\frac{sL^2}2\Bigr)\biggr]
+\frac{\frd0{\mathcal E}{\frac12}P(\mathcal E)}{\sqrt2\pi^\frac32}
\deltaup(L^2),
\end{equation*}
and thus it is obvious that corresponding sufficient condition is
together $\frd\Psi{\frac12}P\ge0$ and
those derived in Sect.~\ref{sec:suf} with $R_0(x)$. With $R(x)$ given
by equation (\ref{eq:rfun}), the preceding discussion in
Sect.~\ref{app:p1} actually extends to $b=0$ thanks to
the non-negativity of $E^\zeta_{p,0}(-z)\ge0$. It follows that
all the corollaries actually hold inclusively for $\beta_1=1$.
[Note the condition $\frd0\Psi\lambda P\ge0$ and
$P(0)=\dotsb=P^{(\flr\lambda-1)}(0)=0$ for
${}^\exists\lambda\ge\frac32-\beta_2\ge\frac32-\beta_1=\frac12$
implies $\frd0\Psi{\frac12}P\ge0$.]

\label{lastpage}
\end{document}